\documentclass[aps,prx,
superscriptaddress, longbibliography]{revtex4-2}
\usepackage{amsmath,amssymb,color,comment,physics,mathtools}
\usepackage{hyperref}
\hypersetup{
    colorlinks=true,
    urlcolor=blue
    }
\usepackage[makeroom]{cancel}
\usepackage{mathrsfs} 
\usepackage{color} 
\usepackage{graphicx}
\usepackage{subcaption}
\usepackage[countmax]{subfloat}
\usepackage[english]{babel}

\usepackage{ulem} 
\usepackage{braket}
 \usepackage{dsfont}
\newcommand{\I}{{\bf I}}

\newcommand{\be}{\begin{eqnarray}}
\newcommand{\ee}{\end{eqnarray}}
\definecolor{mygreen}{rgb}{0,0.5,0}

\newcommand{\ts}{t_{\textrm{sweep}}}
\newcommand{\dep}{\int \frac{d^{3N}P}{(2\pi \hbar_{\textrm{eff}})^{3N}}}

\newcommand{\nn}{\nonumber\\}

\begin{document}
\title{Sampling Rare Conformational Transitions with a Quantum Computer}
\author{Danial Ghamari} 
\affiliation{Department of Physics, University of Trento, Via Sommarive 14, I-38123 Trento, Italy}
\affiliation{INFN-TIFPA, Via Sommarive 14, I-38123 Trento, Italy}
\author{Philipp Hauke} 
\affiliation{INO-CNR BEC Center \& Department of Physics, University of Trento, Via Sommarive 14, I-38123 Trento, Italy}
\author{Roberto Covino}
\email{covino@fias.uni-frankfurt.de}
\affiliation{Frankfurt Institute for Advanced Studies, Ruth-Moufang-Straße 1, D-60438 Frankfurt am Main, Germany}
\author{ Pietro Faccioli}\email{pietro.faccioli@unitn.it}
\affiliation{Department of Physics, University of Trento, Via Sommarive 14, I-38123 Trento, Italy}
\affiliation{INFN-TIFPA, Via Sommarive 14, I-38123 Trento, Italy}
\begin{abstract}
 Spontaneous structural rearrangements  play a central role in the organization and function of complex biomolecular systems. In principle, physics-based computer simulations like Molecular Dynamics (MD) enable us to investigate these thermally activated processes with an atomic level of  resolution. However,  rare conformational transitions are intrinsically hard to investigate with MD, because an exponentially large fraction of computational resources must be invested  to simulate thermal fluctuations in metastable states. Path sampling methods like Transition Path Sampling hold the great promise of focusing the available computational power on sampling the rare stochastic transitions between metastable states. In these approaches, one of the outstanding limitations is to generate paths that visit significantly different regions of the conformational space at a low computational cost. To overcome this issue, we introduce a rigorous approach that integrates a machine learning  algorithm and MD simulations implemented on a classical computer with adiabatic quantum computing. First, using functional integral methods,  we derive a rigorous  low-resolution representation of the system's dynamics, based on a small set of  molecular configurations generated with machine learning.  Then, a quantum annealing machine is employed to explore the transition path ensemble of this low-resolution theory,  without introducing un-physical biasing forces to  steer the system's dynamics.  Using the D-Wave quantum annealer, we validate our scheme by simulating a benchmark conformational transition in a state-of-the-art atomistic description. We show that the quantum computing step generates uncorrelated trajectories, thus facilitating  the sampling of the transition region in configuration space. Our results provide a new paradigm for MD simulations to integrate machine learning and quantum computing.
\end{abstract}
\maketitle

\section*{Introduction}
Molecular dynamics (MD) simulations enable us to investigate the structure and dynamics of molecular systems at high spatial and temporal resolution~\cite{Dror2012}. Despite their large success, MD simulations face the challenge of sampling rare thermally activated re-organizations of complex systems, e.g., conformational changes, folding, and phase transitions~\cite{peters_2017}. Indeed, in a typical simulation, an exponentially large fraction of the computational time will be employed to simulate thermal fluctuations in meta-stable states, rather than sampling the fast transition paths---the stochastic jumps between states---which are rare events~\cite{peters_2017}.  

A wide spectrum of clever enhanced sampling methods have been developed over the last two decades to overcome the rare events sampling problem~\cite{revES}.
Some of these techniques reach a high computational efficiency by introducing history-dependent biasing forces that drive the system out of its thermal equilibrium, thus promoting the escape from meta-stable states (see, e.g.,~\cite{METAD, AMD, TAMD, BFA}). The biasing forces depend on collective variables (CVs), which should encode the essential low-dimensional features of a molecular rare event~\cite{Peters2016}. 
In practice, identifying optimal CVs is a very hard problem, and in realistic conditions sub-optimal CVs will affect the quality of the sampling and the accuracy of the mechanistic understanding emerging from the simulations~\cite{Peters2016}.

\begin{figure}[t!]
\includegraphics[width=1.0\textwidth]{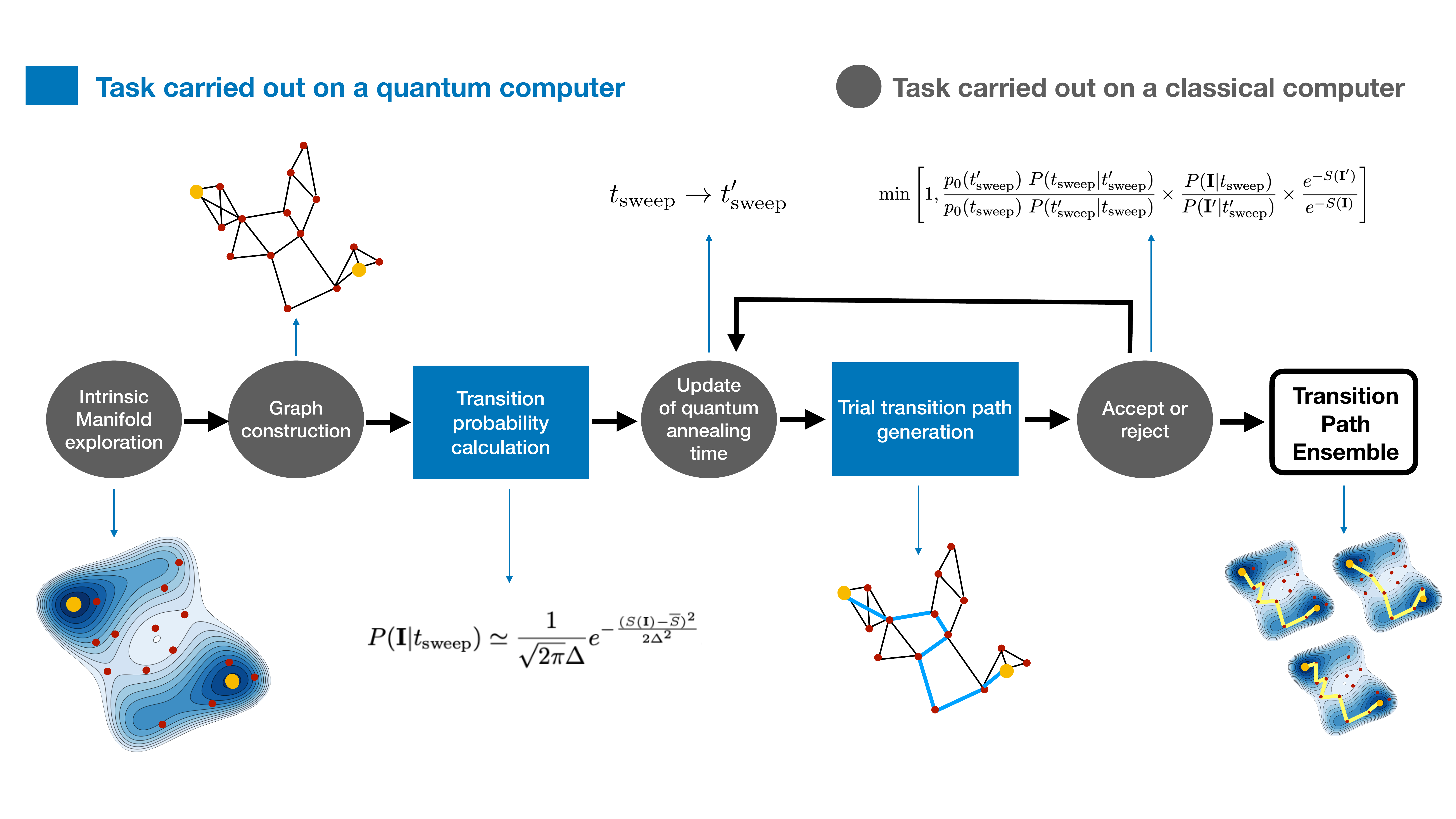}
\caption{Schematic representation of the path sampling scheme introduced in this work, which combines ML and MD performed on a classical computer with QC performed on a quantum annealing machine. Our scheme samples the full transition path ensemble without any use of CVs or unphysical biases. \label{fig:concept}}
\end{figure} 

As an alternative approach, Transition Path Sampling (TPS)~\cite{TPS} is a Markov Chain Monte Carlo scheme that in principle samples the transition path ensemble without involving any biasing force, nor a choice of CVs. 
In TPS, plain MD simulations generate a trial move, i.e., the attempt to generate a new  transition path. For instance, in the so-called shooting move, a new trajectory is initialized from a configuration randomly selected from the last stored transition path~\cite{Jung2017}. Yet, when applied to complex transitions occurring in large configuration spaces with rugged energy landscapes, TPS faces two challenges: efficiently generating viable trial trajectories at an acceptable computational cost and reducing the correlation of generated paths~\cite{Bolhuis2021}. 

Even though promising advancements have recently been made by integrating molecular simulations with machine learning (ML) (see, e.g.,~\cite{CovinoTPS, NoeScience, ParrinelloPNAS, SchneiderPRL, tiwary,No2020,Sidky_2020}), the quest for computationally affordable and accurate enhanced sampling of complex molecular systems remains open. 
In this endeavor, rapid advances in quantum computing provide new opportunities, as is illustrated in the context of quantum chemistry and biology by pioneering applications ~\cite{Q1,Q2,Q3,Q4,Q5,Q6,QDRP,POLYQ}.  
Over the last few years, quantum hardware has grown exponentially both in size and performance~\cite{Scholl_2021,Pogorelov_2021,Ball_2021}, to a point that  it is now realistic to foresee the onset of a tangible quantum advantage in computational problems~\cite{Arute2019,Zhong2020}. It is therefore both important and timely to address the question whether MD, ML, and Quantum Computation (QC) can be   integrated to tackle outstanding challenges of molecular simulations. 

In this work, we integrate MD, ML, and QC to sample the transition path ensemble of thermally activated rare events without involving any unphysical bias or choice of CVs. The salient features of this scheme are illustrated in Fig.~\ref{fig:concept}: First, ML and MD perform a preliminary uncharted exploration of the most visited regions of the configuration space~\cite{IMapD}. These data are used to derive a general coarse-grained description of rare events based on Langevin dynamics. Then, QC on a quantum annealing machine~\cite{DAS_CHAKRABARTI,Das2008,Albash2018,VenegasAndraca2018,Hauke2020} generates transition paths connecting the previously generated configurations. These paths are then accepted or rejected according to a Metropolis criterion implemented on a classical computer, which combines the statistical mechanics of the transition path ensemble with the internal physics of the quantum annealing machine, for which we used the D-Wave machine~\cite{LEAP}. 

We show that the QC step of our sampling scheme generates a new viable and to a good precision uncorrelated transition path at each computing cycle,  and thus can overcome one of the key limitations of path sampling algorithms. 
 
As a first illustrative application, we sample the transition path ensemble of a conformational transition in alanine dipeptide. This benchmark system recapitulates the features of rare events in molecular systems yet is sufficiently small to enable us to encode and run our algorithm on D-Wave. 
Our  results agree well with those obtained by plain MD. From an auto-correlation analysis, we demonstrate that quantum computing generates  uncorrelated trial transition paths at every Monte Carlo step.

The ongoing exponential growth in size and efficiency of quantum computing hardware suggests that, in the future, this same scheme may enable one to investigate transitions that are currently challenging for state-of-the-art classical sampling methods.  
 
  The manuscript is organized as follows. In Section~\ref{theory}, we introduce the general theoretical framework, describe the algorithm used to perform the uncharted exploration of the intrinsic manifold, and our coarse-grained description of reactive processes.
  In Section~\ref{MC}, we discuss the encoding of the path sampling problem on a quantum annealing machine and derive our hybrid Monte Carlo scheme that combines classical and quantum computing. In Section~\ref{application}, we report on our illustrative application to alanine dipeptide. The main results are summarized and discussed in Section~\ref{conclusions}.

\section{Theoretical Setup}  
\label{theory}
In a molecular system at thermal equilibrium, the statistically relevant configurations  accumulate in low-dimensional regions that define the so-called intrinsic manifold. Our path sampling algorithm exploits a recently developed scheme to efficiently explore this intrinsic manifold~\cite{IMapD} (step 1 in Fig.\ref{fig:concept}). Then,  it relies on a coarse-grained representation of the dynamics that is defined directly on this  manifold, based on the configurations generated during the exploration,  to define the input parameters for the quantum annealing part of our algorithm, discussed in Section~\ref{MC} (step 2 in Fig.~\ref{fig:concept}).

\subsection{Uncharted exploration of the intrinsic manifold}

To efficiently explore the intrinsic manifold and sample relevant molecular configurations without any use of CVs or nonphysical bias,  we rely on Intrinsic Map Dynamics (iMapD), a recently developed exploration scheme~\cite{IMapD}. In iMapD, a data-driven manifold learning technique---diffusion maps~\cite{DMAP}---empowers the unbiased MD sampling. Given some local sampling, diffusion maps identify the boundary of the explored configuration space in a low-dimensional representation. New unexplored configurations in the vicinity beyond this boundary are  identified, from which we start a further round of local unbiased sampling. By iterating between these two steps, iMapD rapidly explores the relevant parts of the configuration space. In the Appendix, we summarise the key aspects of the theory and implementation of iMapD.

At the end of the exploration, iMapD yields a data set of configurations $\mathcal{C}=\{Q_k\}_{k=1,\ldots, \nu}$, that by construction lie on the intrinsic manifold, and that were obtained at a much lower computational cost than by using equilibrium MD~\cite{IMapD}.

\subsection{Coarse-grained representation of the dynamics on the intrinsic manifold}

Once the data set  $\mathcal{C}$ is established, we can use it to build a coarse-grained representation of the dynamics, defined directly on the intrinsic manifold explored with iMapD.
 
The sparse configuration data set  generated in the uncharted exploration defines a partition of the intrinsic manifold in finite sub-regions. The $i$-th region is identified with the neighborhood  of configuration $Q_i\in \mathcal{C}$. For example, in a Voronoi tassellation, $Q_i$  would represent the center of a  cell whose boundaries lie midway to the neighboring sampled configurations (see Fig.~\ref{fig:coarse-grained}). 
 
The spatial resolution scale $\sigma$ of this coarse-grained representation of the molecular dynamics is set by the average distance between neighboring configurations in the data set $\mathcal{C}$. The  temporal resolution $\Delta t$ is estimated by the average time the  system takes to diffuse across neighboring regions. At this level of coarse-graining, transition pathways correspond to ordered sequences of visited sub-regions  (see, e.g., the yellow regions in Fig.~\ref{fig:coarse-grained}). Therefore, a transition path  can be specified by an integer vector $\I = (i_1, \ldots, i_{N_I})$, where $i_k$ is the label pointing at the neighborhood of the configuration $Q_{i_k}$, which is visited at the  $k$-th time step. 

In what follows, we develop a rigorous statistical mechanical formalism that enables us to describe the coarse-grained dynamics on the intrinsic manifold. To this end, we employ a powerful path-integral formalism combined with so-called regularization and renormalization procedures that were originally developed in the framework of nuclear and subnuclear physics (see~\cite{Howto} for an enlightening pedagogical introduction). 

\subsubsection*{ Path integral formulation of stochastic dynamics} The goal of this subsection is to derive an expression for the probability of  arbitrary coarse-grained paths $\I$  on the intrinsic manifold, in the form
\begin{eqnarray}\label{effT}
P(\I) &\propto& e^{-S({\bf I})},
\ee 
where the functional $S(\I)$
is called the effective action of the coarse-grained path $\I$. 
\begin{figure}[t!]
\includegraphics[width=.8\textwidth]{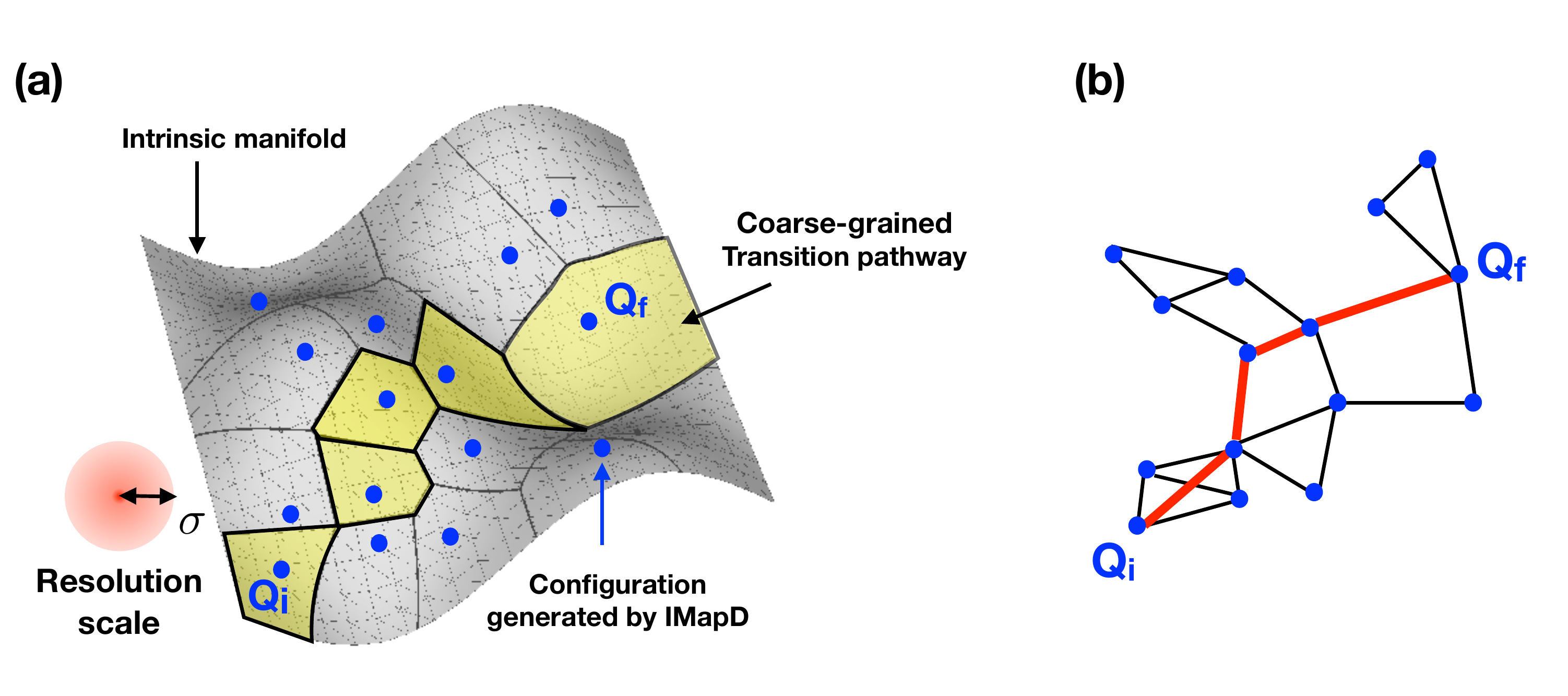}
	\caption{(a) Illustration of the coarse-grained representation of the Langevin dynamics. Dots represents configurations generated using iMapD, which lie on the intrinsic manifold. Each such configuration is regarded as a representative element of its Voronoi cell. Coarse-grained trajectories are identified by sequences of Voronoi cells, $I$. A typical transition path is highlighted in yellow. The intrinsic resolution scale $\sigma$ of the effective theory is set by the average distance between the configurations generated by iMapD.  (b) Graph representation of the coarse-grained Langevin dynamics. The red line denotes the transition path highlighted in yellow in panel (a). 
	    	\label{fig:coarse-grained}
	}
\end{figure} 

We assume the system obeys a Langevin dynamics in the overdamped limit:
\be\label{EQ:langevin}
\dot Q(t) = -\frac{1}{m \gamma}\nabla U(Q) + \eta(t)\,,
\ee
where $Q$ is a point in the molecular configuration space, $m$ and $\gamma$ are the atomic mass an viscosity (assumed for simplicity to be uniform throughout the system), $U(Q)$ is the potential energy, and $\eta(t)$ is a white noise of null average obeying the fluctuation-dissipation relationship
\be
\langle \eta(t) \cdot \eta(0) \rangle = 6 D N_a  \delta(t),\,
\ee
where $N_a$ is the number of atoms and $D=k_{\textrm{B}}T/m\gamma$ is the diffusion coefficient.
The probability of observing a system described by Eq.~(\ref{EQ:langevin}) in a given configuration  at a given time obeys the Fokker-Planck equation:
\be
\frac{\partial}{\partial t} P(Q,t) = D \left(\nabla^2 + \frac{1}{k_\mathrm{B}T} \nabla U(Q)\cdot  \nabla + \frac{1}{k_\mathrm{B}T}\nabla^2 U(Q)  \right) P(Q,t)\,.
\ee
The solution  of this  equation can be conveniently expressed in path-integral form~\cite{DRP0, DRP1, Donniach},
\begin{equation}\label{Pcond0}
    P(Q_f, t_f|Q_i, t_i) =\int_{Q_i}^{Q_f} \mathcal{D}Q~e^{-S_{\mathrm{OM}}[Q]},
\end{equation}
where $S_{\mathrm{OM}}[Q]=\frac{1}{4D} \int_{t_i}^{t_f} d\tau \left(\dot Q + \frac{D}{k_\mathrm{B}T}\nabla U[Q(\tau)]\right)^2$ is the so-called Onsager--Machlup functional~\cite{OM}.
Equation~(\ref{Pcond0}) can be re-written in an equivalent form that resembles that of an imaginary-time ``quantum" Feynman propagator~\cite{Caroli,Autieri} 
\be\label{Pcond1}
P(Q_f, t|Q_i, t_i) &=& e^{-\frac{1}{2 k_\mathrm{B}T} \left( U(Q_f)-U(Q_i)\right)} ~K(Q_f, t| Q_i,t_i), \\
\label{Pcond2}
K(Q_f, t|Q_i, t_i)  
&=& \int_{Q_i}^{Q_f}\mathcal{D}Q~e^{-\frac{1}{\hbar_{\mathrm{eff}}} \int_{t_i}^{t_f} d\tau \left(\frac{\dot Q^2}{2}m + V_{\mathrm{eff}}[Q(\tau)] \right)}, 
\ee
where $m$ is a uniform atomic mass  and we have introduced an effective ``Dirac's constant" $\hbar_{\textrm{eff}}= \frac{2 k_\mathrm{B}T}{\gamma}$ that  controls the amount of thermal fluctuations in the system, and
\be\label{Veff}
V_{\textrm{eff}}(Q) = \frac{1}{2 m \gamma^2}\left[|\nabla U(Q)|^2 - \hbar_{\textrm{eff}}  \gamma \nabla^2 U(Q)\right]
\ee
is called the  effective potential. 
The pre-factor $e^{-\frac{1}{2 k_BT} \left( U(Q_f)-U(Q_i)\right)}$ in Eq.~(\ref{Pcond1}) does not affect the relative statistical weight of transition paths sharing identical boundary conditions. 
 
To build a  coarse-grained representation of the molecular dynamics with a spatio-temporal resolution set by the cut-off scales $\sigma$ and $\Delta t$, it is convenient to adopt the mathematical formalism of quantum mechanics and re-express the path-integral $K(Q_f, t|Q_i, t_i)$ as a matrix element:
\be\label{Kbraket}
K(Q_f, t|Q_i, t_i) =\langle Q_f| e^{-\frac{1}{\hbar_{\textrm{eff}}} \hat H_{\textrm{eff}} (t-t_i)}|Q_i\rangle, 
\ee
where $\hat H_{\textrm{eff}} = -\frac{\hbar_{\textrm{eff}}^2}{2 m} \nabla^2 + V_{\textrm{eff}}(Q)$ is the corresponding effective  Hamiltonian. 
This description of stochastic dynamics is formulated on the Hilbert space spanned by the molecular configuration eigenstates $|Q_i\rangle$. 

 \subsubsection*{Regularization}  The advantage of this Quantum Mechanical formulation is that the resolution power of our theory can be lowered by removing the large momentum states from the Hilbert space.   A practical way to do so is to introduce a Gaussian cut-off in all momentum integrals:
\be\label{reg}
\dep \to \dep ~e^{-\frac{P^2 \sigma^2}{2 \hbar^2_{\textrm{eff}}}}\,. 
\ee 
In the quantum mechanical terminology, this filtering procedure is usually referred to as ``regularization" step of the renormalization procedure.  
As a first illustrative example of regularization, we consider the inner product between position eigenstates:
\be \langle Q|Q'\rangle &=& \dep  ~e^{\frac{i}{\hbar_{\textrm{eff}}} P\cdot  (Q'-Q)} = \delta(Q-Q').
\ee
Its regularization yields
\be
\langle Q| Q'\rangle_{\mathrm{reg}} &=&  \dep ~e^{\frac{i}{\hbar_{\textrm{eff}}} P\cdot  (Q'-Q)}  e^{-\frac{P^2 \sigma^2}{2 \hbar_{\textrm{eff}}^2}}\nn   &=& \frac{1}{(\sqrt{2 \pi} \sigma)^{3N_a} } e^{-\frac{(Q'-Q)^2}{2 \sigma^2}} \equiv \delta_\sigma(Q-Q').
\ee
Note that $\delta_\sigma(Q-Q')$  provides an effective representation of Dirac's delta function, smeared to the desired spatial  resolution scale  $\sigma$. 

The same regularization prescription can be applied to remove short-distance details from the Feynman propagator Eq.~(\ref{Kbraket}).  The starting point consists in applying the standard Trotter decomposition to obtain
\begin{equation}
K_{\textrm{reg}}(Q_f,t|Q_i) =  \int \mathrm{d}Q_0 \ldots dQ_{N_t}~ \prod_{n=0}^{N_t-1}~\left[\langle Q_{n+1}| e^{-\frac{1}{\hbar_{\textrm{eff}}} \hat H_{\textrm{eff}} \Delta t}|Q_{n}\rangle_{\textrm{reg}}\right] \,\delta_\sigma(Q_{N_t}-Q_f)\,\delta_\sigma(Q_{0}-Q_i).
\end{equation}
In  Appendix~\ref{appendixa}, we explicitly show that
\be\label{Kreg}
\langle Q_{n+1}| e^{-\frac{\Delta t}{\hbar_{\textrm{eff}}}\hat H_{\textrm{eff}}}|Q_{n}\rangle = \mathcal{N} ~e^{-\frac{1}{\hbar_{\textrm{eff}}}\left[ C_{\textrm{T}} \frac{m}{2} \left(\frac{Q_{n+1}-Q_{n}}{\Delta t}\right)^2   +  V_{\textrm{eff}}^{\textrm{reg}}(Q_n)\right]\Delta t},
\ee
where 
\be\label{CT}
C_{\textrm{T}} = \left(1 +  \frac{  m \sigma^2}{  \hbar_{\textrm{eff}}\Delta t}\right)^{-1}
\ee
 and $V^{\textrm{reg}}_{\textrm{eff}}(Q)$ is the regularized effective potential, defined in Eq.~(\ref{veffR}).
 
Even though the direct evaluation of $V^{\textrm{reg}}_{\textrm{eff}}(Q)$ is computationally challenging, we can consider approximations to make it feasible in practice. 
For example,  we can smear out its short-distance structure  performing a self-averaging over the values of the effective potential evaluated on groups of structurally close configurations generated during the IMapD exploration. 
However, regularizing  the effective potential $V_{\textrm{eff}}(Q)$ is  not sufficient to define the correct path probability in the coarse-grained theory.
To see this, we consider the probability $P_R(Q)$ for a system initially at $Q$ to remain within an infinitesimal volume $dV$ after a short time $\Delta t$. From the path-integral expression, Eq.~(\ref{Pcond2}), it follows that $ P_R(Q) \propto dV e^{-\frac{V_{\textrm{eff}}(Q)\Delta t}{\hbar_{\textrm{eff}}}}$. 
Thus, $V_{\textrm{eff}}(Q)/\hbar_{\textrm{eff}}$ is related to  the rate of escape from an infinitesimal volume centered around $Q$.
Similarly, the effective potential in the coarse-grained theory, $V_{\textrm{eff}}^{\textrm{cg}}(Q_i)$ -- which determines the path probability -- should be related to the rate of escape from the finite region identified with the neighborhood of the data point $Q_i\in \mathcal{C}$. 
We account for this effect at the phenomenological level\footnote{We note that, in principle,  a fully rigorous approach to computing $V_{\textrm{eff}}^{\textrm{cg}}(Q)$ would involve the Renormalization Group-based effective theory construction, as discussed in \cite{Howto}. In practice, however, this program would be extremely challenging to implement for realistic molecular systems.}, by re-scaling the regularized effective potential by a factor $C_{\textrm{V}}$:
\be
V_{\textrm{eff}}^{\textrm{cg}}(Q_i) \simeq C_{\textrm{V}} V_{\textrm{eff}}^{\textrm{reg}}(Q_i).
\ee 

After  multiplying all elementary propagation terms and restoring the continuum notation, the final expression for the regularized Feynman propagator of the coarse-grained theory reads
\be\label{Kreg_final}
K_{\textrm{reg}}(Q_f, t|Q_i, t_i)  
&=& \int_{Q_i}^{Q_f}\mathcal{D}Q~e^{-\frac{1}{\hbar_{\textrm{eff}}} \int_{t_i}^{t_f} d\tau \left(C_{\textrm{T}} \frac{\dot Q^2}{2}m + C_{\textrm{V}} V^{reg}_{\textrm{eff}}[Q(\tau)] \right)}.
\ee

 \subsubsection*{Renormalization}  The dimensionless renormalization constants $C_{\textrm{T}}$ and $C_{\textrm{V}}$  implicitly depend on our choice of cut-off scales $\sigma$ and $\Delta t$. They could be determined by matching against microscopic calculations.  However, for the purpose of the present work, an order-of-magnitude estimate suffices. 
To estimate $C_{\textrm{T}}$, we recall that, to leading order in $\Delta t$, our spatial and temporal cut-off scales are related by an approximate Einstein's relation, $  N_a D \Delta t \sim \sigma^2$. Using $\hbar_{\textrm{eff}}= 2 k_\mathrm{B}T/\gamma$ and the Einstein's relation in Eq.~(\ref{CT}), we obtain  $C_{\textrm{T}} \sim \frac{1}{N_a}$. Since $(Q_i-Q_j)^2 \sim  N_a D \Delta t$, the effective kinetic energy term entering in the exponent of the right-hand-side of Eq.~(\ref{Kreg}) is of order 1 and does not scale with the number of atoms in the system. This is important to ensure that the relative probability of different coarse-grained path remains finite. To estimate $C_{\textrm{V}}$, we recall that the effective potential of the coarse-grained theory $V_{\textrm{eff}}^{cg}(Q_i)$ is related to the rate of escape from the region associated with the neighborhood of $Q_i$ in the intrinsic manifold.

Since $\Delta t$ is the typical resolution scale of the coarse-grained theory,  $C_{\textrm{V}} V_{\textrm{eff}}^{\textrm{reg}}(Q) \Delta t$ in the right-hand-side of Eq.~(\ref{Kreg}) must be of order~1. 

\subsubsection*{Path probability in the graph representation} Equation~(\ref{Kreg_final}) leads to a closed  expression for the coarse-grained effective path action entering Eq.~(\ref{effT}):
\be
S(\I) &=& \sum_{k} w_{i_{k+1} i_k}, 
\ee
where
\be
\label{wijdef}
w_{i j} &\equiv&  C_{\textrm{T}} \frac{(Q_{i}-Q_{j})^2}{4 D  \Delta t}  + C_{\textrm{V}} V_{\textrm{eff}}^{\textrm{s}}(Q_{i})\Delta t.
\ee
The exponent $e^{-w_{ij}}$ controls the probability of observing a transition between the regions $I_i$ and $I_j$ of the intrinsic manifold, in an elementary (coarse-grained) time step $\Delta t$. Since both terms at the exponent of Eq.~(\ref{wijdef}) are of order 1, the probabilities of different paths remain finite and comparable. Further details about explicit evaluation of the weights $w_{ij}$ in a realistic application are reported in the Appendix~\ref{appendixd}. Here, we only emphasize that we  fix the renormalization parameters $C_{\textrm{T}}$ and $C_{\textrm{V}}$ by imposing the condition that, following our above estimates, all $w_{ij}$ should be of order~1. In practice, this is achieved by dividing each weight by the largest edge weight in the graph, i.e., $w_{ij}\to w_{ij}^R \equiv w_{ij}/w_{max}$. 

\section{ Transition Path Sampling with a quantum annealing machine.} 
\label{MC}

Designing sampling algorithms exploiting quantum annealers has become a highly active research field, especially with applications to machine learning~\cite{adachi2015application,Chancellor_2016,Benedetti_2017,Winci_2020,Sieberer_2018}. We leverage on this development by integrating sampling via a quantum annealer into our classical--hybrid scheme, in order to generate realistic ensembles of coarse-grained  transition pathways $\I$. To this end, we need to sample from the path distribution $e^{-S(\I)}$. 
In principle, conventional stochastic algorithms on a classical computer could serve this purpose. However, their computational cost grows very rapidly with the number of configurations in the data set $\mathcal{C}$.  Ultimately, classical Markov chain Monte Carlo path sampling algorithms are typically limited by  long auto-correlation times in the chain. As we will show below, quantum computers can overcome this limitation: in our approach each Monte Carlo step performed on D-Wave can generate a new uncorrelated transition path. Note that we do not require a fully fair sampling of the space of possible paths, which is one of the challenges in quantum-annealer based sampling~\cite{K_nz_2019,Yamamoto_2020,kumar2020achieving}. Employing a suitable reweighting procedure, it is sufficient for our algorithm if the exploration of the accessible space is sufficiently broad.

\subsection{Quantum encoding of the transition path sampling problem.}

The first step to  derive our path sampling algorithm consists in  introducing a graph representation of the path probability density defined in Eq.~(\ref{effT}). We identify each configuration in the data set $\mathcal{C}$  with a node in the  graph and define the  topology of the graph so to ensure that connected neighboring nodes represent configurations that are both structurally and kinetically close (in Appendix~\ref{appendixd}, we provide further details on how we enforce this condition in the application to alanine dipeptide).
The weights $w_{i j}$ of the edges in the graph are defined according to Eq.~(\ref{wijdef}), thus ensuring that the sum of the weights along a given path $\I$ on the graph yields the path functional $S(\I)$ entering Eq.~(\ref{effT}).

 The undirected graph representation enables us to map the sampling problem to a quantum annealing one. To this end,  we introduce two sets of binary variables, $\Gamma^{(1)}_i$ and $\Gamma^{(2)}_{ij}$, where $i$ and $j$ run over the $\nu$ vertexes in the graph.  If $\Gamma^{(1)}_i=1$ ($\Gamma^{(1)}_i=0$), then the $i$-th node is (is not) visited by the transition path on the graph (see red line in Fig.~\ref{fig:coarse-grained}(b)). $\Gamma^{(2)}_{ij}$ is always 0 if the $i$ and $j$ are not adjacent in the graph. If $i$ and $j$ are adjacent, then $\Gamma^{(2)}_{ij}=1$ when the path contains the $i\to j$ or $j \to i$ transition. We are specifically interested  in configurations of the binary variables in which the set of non-vanishing  entries of $\Gamma^{(1)}_i$ and $\Gamma^{(2)}_{ij}$   form a topologically connected path, i.e., a continuous line starting from the given initial node and terminating in the chosen final node.

To sample path configurations according to $e^{-S({\bf I})}$, let us  consider the following classical Hamiltonian of the binary variables:
\begin{eqnarray}
\label{HQUBO}
H =  \alpha H_{\textrm{C}} + H_{\textrm{T}}.
\end{eqnarray} 
$H_{\textrm{C}}$ is the constraint Hamiltonian, a positive-definite function that is zero only if the entries of the binary variables satisfy the path topology, $H_{\textrm{C}}(\Gamma^{(1)}, \Gamma^{(2)})=0$.
This condition can be fulfilled by choosing~\cite{shortest_path_on_network}
\be
H_{\textrm{C}}= H_{\textrm{s}}+H_{\textrm{t}}+H_\textrm{r},
\ee
where 
\be
H_{\textrm{s}}=& -\left(\Gamma^{(1)}_s\right)^2+\left(\Gamma^{(1)}_s-\sum_i\Gamma^{(2)}_{s i}\right)^2\,,\\
H_{\textrm{t}}=&- \left(\Gamma^{(1)}_t\right)^2+\left(\Gamma^{(1)}_t-\sum_i\Gamma^{(2)}_{t i}\right)^2\,,\\
H_{\textrm{r}}=& \sum_{j\ne s,t} \left(2\Gamma^{(1)}_j-\sum_i\Gamma^{(2)}_{j i}\right)^2\,.
\ee
In this formulation, $H_\textrm{s}$ and $H_\textrm{t}$ introduce the condition that the path should start from the initial node $s$ and end in the final node $t$, while $H_\textrm{r}$ imposes the flux conservation at the remaining nodes.

$H_\textrm{T}(\Gamma^{(2)})$ is the so-called target function, which reads  $H_\textrm{T}= \sum_{ij} w_{ij}~\Gamma^{(2)}_{ij}$ and whose physical interpretation is as follows. By definition, $H_\textrm{T}$ yields the path action $S$  whenever the configuration of the tensors $\Gamma^{(1)}$ and $\Gamma^{(2)}$ satisfy a path topology, that is, if $H_\textrm{C}( \Gamma^{(1)},\Gamma^{(2)})=0$ and $\I=\I(\Gamma^{(1)}, \Gamma^{(2)})$ is the corresponding path, then  $H_\textrm{T}(\Gamma^{(2)})=S(\I)$. The parameter $\alpha$ in Eq.~(\ref{HQUBO}) controls the relative strength of the constraint Hamiltonian, $H_\textrm{C}$. For  $\alpha\gg 1$, all binary variables'  configurations  that violate  the path topology correspond to very high excitations and are thus excluded from the sampling in the low-energy states that the quantum annealer performs.

Now, we are finally in a condition to tackle the problem of how to use a quantum annealer to sample path configurations with probability distribution $\propto e^{-S(\I)}$. First, we use this machine to  generate path-like binary variables' configurations according to a probability distribution that has a finite overlap with $e^{-S(\I)}$. Then, a classical machine accepts or rejects the proposal, thus restoring the correct path probability distribution $e^{-S(\I)}$ (see Fig.~\ref{fig:concept}).   

To implement this scheme, we switch to a generalized Ising Model formulation of our classical Hamiltonian, by means of a change of variables:  $\sigma_i^z=2 \Gamma^{(1)}_i-1$,  $\sigma_{ij}^z= 2 \Gamma^{(2)}_{ij}-1$. 
Then, we promote the classical  Eq.~(\ref{HQUBO}) to a quantum mechanical Hamiltonian $\hat H$, by substituting the classical Ising variables  with Pauli $z$ operators of a spin 1/2 algebra. Finally, the spin $1/2$ states are encoded in the qubits of D-Wave. 

In a standard quantum annealing process, the qubits are initialized in the ground-state of an easily solvable Hamiltonian~\cite{DAS_CHAKRABARTI,Das2008,Albash2018,VenegasAndraca2018,Hauke2020}, in our case 
\be
\hat H_\textrm{in}= -h_\mathrm{x} \left(\sum_{i}\hat \sigma^x_i -  \sum_{ij} \hat \sigma_{ij}^x\right),
\ee
where $h_\mathrm{x}$ is an arbitrary real constant.  Then, the system is subjected to a time-dependent Hamiltonian 
\be
\hat H(t) = A(t)~\hat H_{\textrm{in}} + B(t)~\hat H,\ee
with scheduling functions $A(t)$ and $B(t)$. These are chosen such that initially $A(0) = 1$ and $B(0) = 0$, while at the end of the protocol, i.e., at $t=t_{\textrm{sweep}}$, one has $A(t_{\textrm{sweep}}) = 0$ and $B(t_{\textrm{sweep}}) = 1$. 

The spectrum of the quantum Hamiltonian $\hat H$ comprises all possible energy states of the classical Hamiltonian $H$ defined in Eq.~(\ref{HQUBO}). Therefore, as long as $\alpha\gg 1$, the  low-lying eigenstates of  $\hat H$ represent path-like configurations $\I$ and their eigenvalues coincide with the path action $S(\I)$. For a closed system, the adiabatic theorem implies that if the sweep is performed sufficiently slowly as compared to the minimal energy gap $\Delta E$, i.e., for $t_{\textrm{sweep}}\gg \hbar/\Delta E$, then the system remains in its instantaneous ground-state, thus reaching the lowest energy solution at the end of the sweeping process.
In this ideal condition, the annealing process would systematically return the least action path $\bar{\I}$~\cite{QDRP}. Since the path probability in Langevin dynamics is given by $\sim e^{-S(\I)}$, the least action path $\bar{\I}$  corresponds to the most probable transition path.

In realistic conditions, the probability of landing onto the ground-state remains $<1$, even in the limit of very long sweeping times. This is because  the coupling of the quantum annealing device to its environment induces decoherence and thermal relaxation~\cite{sweep1}. It has been suggested that this coupling can be exploited to sample classical Boltzmann distributions~\cite{samp1, samp2}. However, 
in practice, the sampling can only be performed at some rescaled temperature that is very difficult to estimate {\it a priori}~\cite{samp2}. 
The reason is that, if the coupling $A(t)$ of the initial Hamiltonian $H_{\textrm{in}}$ decays sufficiently fast, the thermal relaxation time may grow longer than the  sweeping time $\ts$, and the relaxation process freezes at some time $t_f<\ts$.  In this case, the distribution of final energy states would be close to a modified Boltzmann distribution  $e^{-B(t_f) S({\bf I})}$, where $t_f$ is the freezing time. It should be emphasized, however, that the  existing quantum annealing machines such as D-Wave, are very often employed in hybrid  optimization schemes that combine classical and quantum annealing. In this case,  we do not expect the path probability should correspond to a Boltzmann distribution. In the following, we only assume that there exists a regime of sweeping times for which the distribution of the paths generated by multiple hybrid energy minimization has a finite overlap with $e^{-S(\I)}$.

A classical computer controlling a Metropolis scheme can exploit this overlap to yield the correct  sampling of $e^{-S(\I)}$ (Fig.~\ref{fig:concept}). 
In general, this can be achieved by imposing the detailed balance condition $e^{-S(\I)}  T( \I'|\I) = e^{-S(\I')} T(\I|\I')$,
where $T(\I'| \I)$ is the transition probability from the path ${\bf I}$ to the path ${\bf I}'$ in the underlying stochastic process. We choose to generalize this dynamics  to enable also the sweeping time $\ts$ to vary along the Markov chain. We do so to ensure that D-Wave is mostly performing sweeps with a duration $\ts\sim t_0$, where $t_0$ is a tunable parameter representing a reasonable compromise between accuracy (slow sweeping) and efficiency (low consumption of quantum computing time).
Upon enlarging the configuration space of the Monte Carlo dynamics to include $\ts$,  the new detailed balance condition reads $\rho(\I, \ts)~ T(\ts', \I'|\ts, \I) = \rho(\I', \ts')~T(\ts, {\bf I}|(\ts', \I')$, where $\rho(\I, \ts) $ is the new equilibrium distribution. Our Monte Carlo dynamics must be defined in such a way to ensure that the equilibrium distribution is
\be
\rho(\ts, \I) = p_0(\ts)\times e^{-S(\I)},
\ee
where $p_0(\ts)$ is some arbitrary equilibrium distribution of the sweeping time, centered around $t_0$.  
Following the standard procedure to obtain the Metropolis acceptance/rejection criterium, we write  the transition probability  as a product of a trial move probability $\tau(\I', \ts'|\I, \ts)$ and a corresponding acceptance probability $a(\I', \ts'|\I, \ts)$. 
Since the sweeping time is allowed to vary along the chain, we factorize the trial move probability as
\begin{equation}
\tau(\I', \ts'| \I ,\ts) = P(\ts'|\ts) ~P(\I'|\ts'),
\end{equation}
where $P(\ts'| \ts)$ is the probability for the sweeping time to go from $\ts$ to $\ts'$ in a Monte Carlo step, while $P(\I|\ts)$ is the probability that a quantum annealing calculation lasting a time $\ts$ yields the path  $\I$. 
Combining all terms together, we obtain the following Metropolis acceptance rule:
\begin{eqnarray}\label{detbal}
\textrm{min} \left[1, \frac{p_0(\ts')~P(\ts| \ts')}{p_0(\ts)~P(\ts'| \ts)} \times \frac{P(\I|t_{\textrm{sweep}})}{P(\I'|t'_{\textrm{sweep}})}\times \frac{e^{-S(\I')}}{e^{-S(\I)}} \right].
\end{eqnarray}

In particular, in our simulations we chose to update $\ts$ according to a Brownian dynamics with a harmonic drift term: 
\be
\label{evolts}
 t_{\textrm{sweep}}^{i+1} = t_{\textrm{sweep}}^i  -\delta t k (\ts-t_0) + \sqrt{2 \delta t} \xi^i,
\ee
 where $\xi_i$ is a Gaussian distributed random variable of null mean and unitary variance and $\delta t$ is an incremental sweeping time change.

The conditional probability $P(\I|\ts)$ in Eq.~(\ref{detbal}) depends on the details of the quantum annealing machine and of the specific optimization algorithm. In general, computing $P(\I| \ts)$ from a theoretical model of the annealing process can be very challenging. We overcome this problem and show how to estimate $P(\I|\ts)$  by performing a moderate number of annealing processes, for each value of $t_{\textrm{sweep}}$. 
The spectrum of the target quantum Hamiltonian $\hat H$ is expected  to be non-degenerate, since the weights in the graph  $w_{ij}$ are in general all different. In addition, for large values of the parameter $\alpha$ in Eq.~(\ref{HQUBO}), all low-lying states satisfy the constraints set by $H_\textrm{C}$, and thus correspond to path-like configurations $\I$. 
Therefore, each low-lying eigenvalue $E$ of the quantum Hamiltonian $\hat H$ corresponds to the action of a single  path, $E=S(\I)$.
Then, $P(\I|\ts)$ can be directly inferred from a frequency histogram of the energies $E$ obtained at the end of multiple annealing processes performed at fixed $\ts$, i.e., $P(\I|\ts)=P(E| \ts)$. To minimize the consumption of quantum computing time, we can estimate $P(E| \ts)$ by the lowest-order cumulant expansion as 
\begin{eqnarray}\label{cumul}
P(\I|\ts) \simeq P(E|\ts) \simeq \frac{1}{\sqrt{2 \pi} \Delta}
e^{-\frac{(E(\I)-\overline{E})^2}{2 \Delta^2}},
\end{eqnarray}
where $\overline{E}$ and $\Delta$, respectively, are the average and the standard deviation of the energy  obtained by many annealing processes  at fixed sweeping time $\ts$.

\section{ Application to a molecular benchmark system.}
\label{application}
To illustrate our hybrid  classical/quantum Monte Carlo scheme sketched in Fig.~\ref{fig:concept}, we apply it to simulate
the $C_5\to\alpha_R$ transition of alanine dipeptide. This standard benchmark system is sufficiently small to enable us to carry out the quantum computing calculations on the D-Wave. 

First, we use iMapD and our effective Langevin theory to construct the graph representation of the dynamics on the intrinsic manifold of this peptide (details on the implementation of iMapD and the calculation of the weights $w_{ij}$ for this molecular system are provided in the  Appendix~\ref{appendixc} and  Appendix~\ref{appendixd}). The results are shown in the Ramachandran plot reported in Fig.~\ref{fig:network}.  The contour lines in the background represent the free energy surface, calculated from a frequency histogram of  $1~\mu$s of equilibrium MD at $T=300~$K, generated using OpenMM~\cite{OpenMM}, in the AMBER99SB force field with explicit TIP3P water~\cite{forcefield}.
\begin{figure}[t!]
    \includegraphics[width=.48\textwidth]{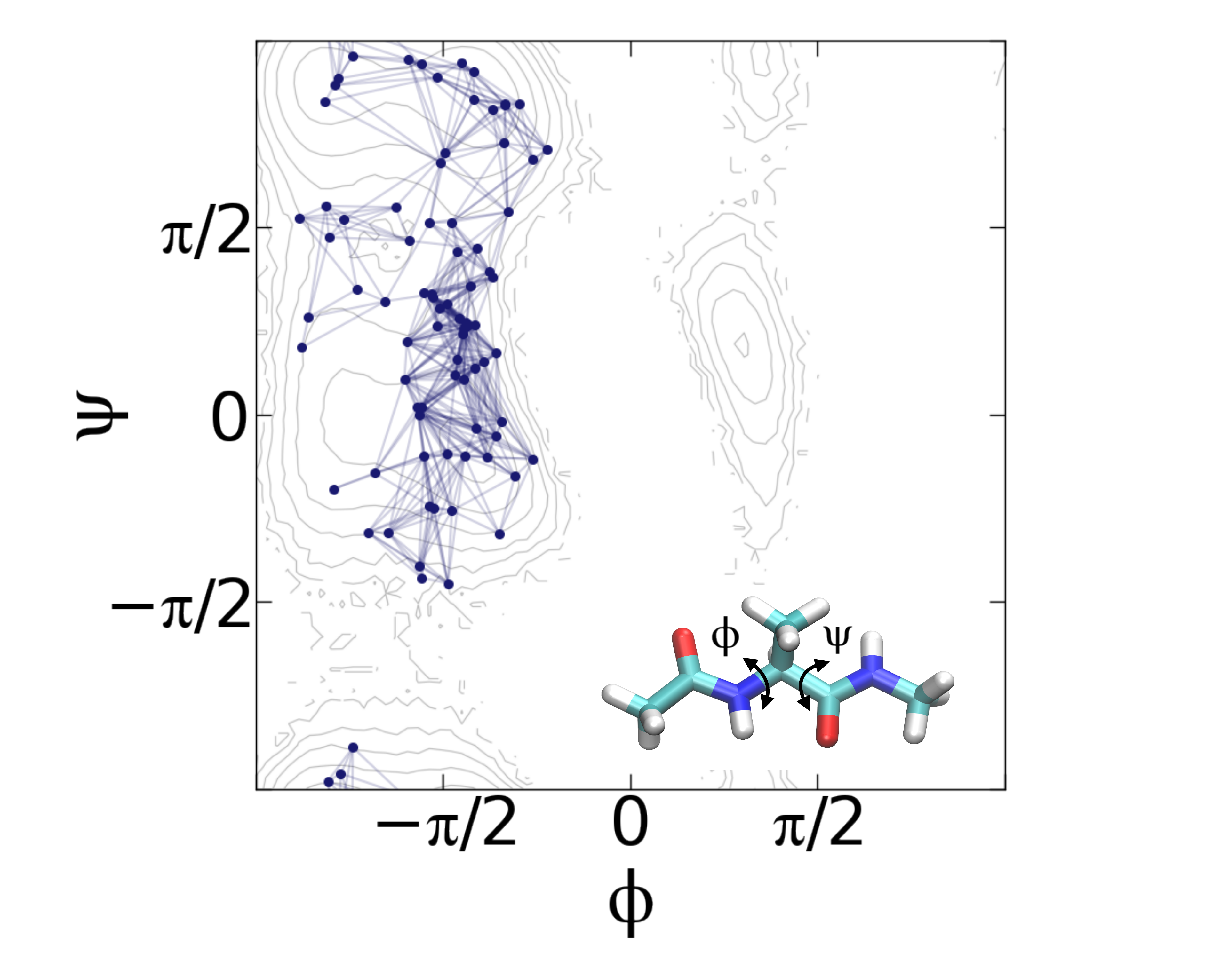}
	\caption{Network of configurations plotted on the Ramachandran plot defined by the dihedral angles $\phi$ and $\psi$ of alanine dipeptide. Nodes correspond to molecular configurations derived using iMapD exploration, and edges connect configurations that are kinetically and structurally close. The number of nodes and edges are $\nu=83$ and $|\mathcal{E}|=495$, respectively. The background shows isolines of the free energy estimated from an equilibrium MD simulation, plotted every 3 kJ/mol.}
		\label{fig:network}
\end{figure} 
\begin{figure}[t!]
    \includegraphics[width=0.5\textwidth]{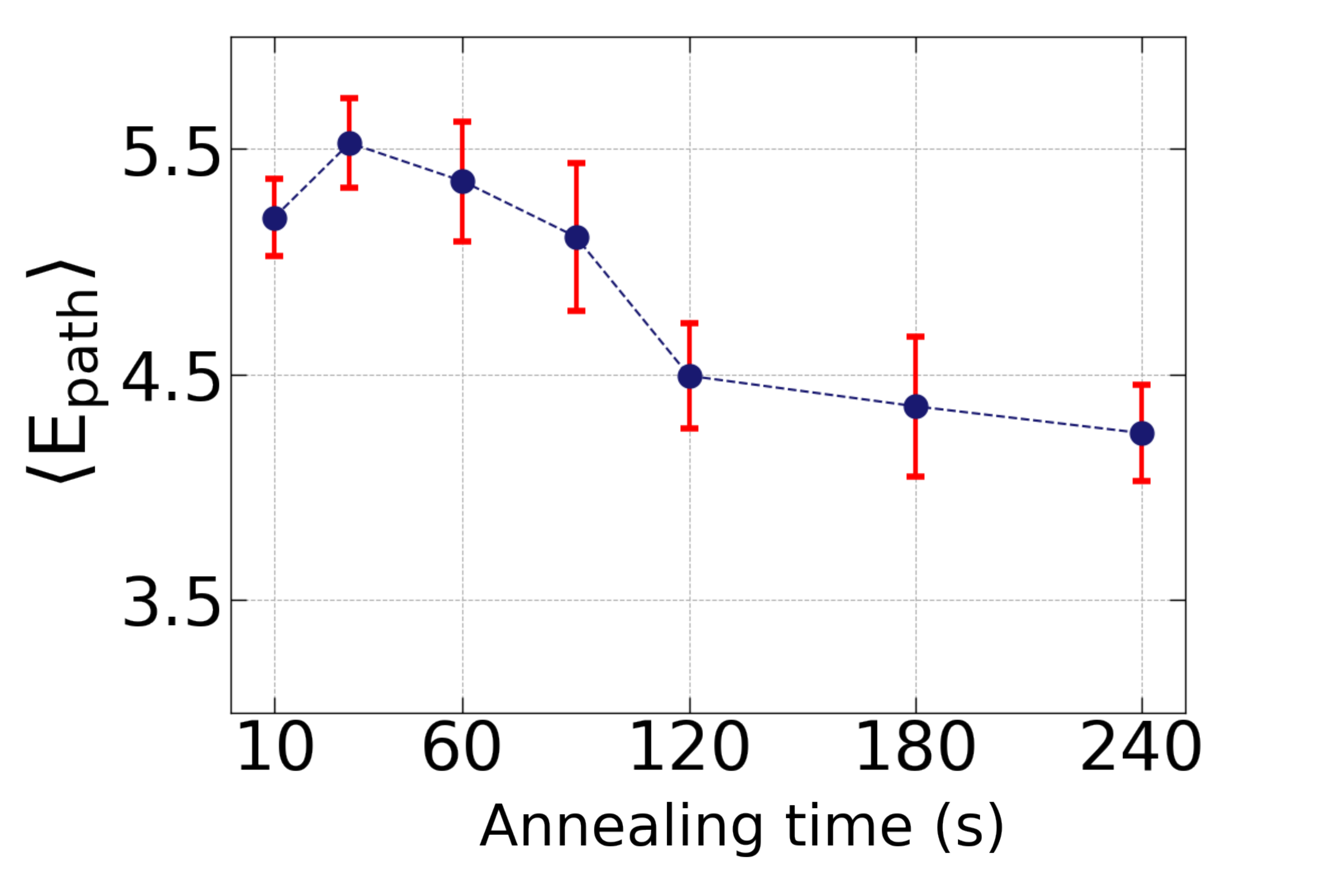}
	\caption{Average value and standard error of the mean of the energy $E$ obtained
	by multiple quantum annealing processes at fixed values of $\ts$. These results are used to estimate $P(E|\ts)$ to lowest order in the cumulant expansion approximation (\ref{table:table1}).\label{Pcondest}}
\end{figure}
The spatial resolution of our effective theory is determined by the number of configurations $\nu$ we keep to generate a sparse graph. With this choice, the average RMSD distance between neighbouring configurations in our network is $\delta_{\mathrm{RMSD}}\simeq 0.5$ nm. Then, $\sigma \simeq \delta_{\mathrm{RMSD}} \sqrt{N_a}$, where $N_a=22$ is the number of atoms in our molecule. The time scale $\Delta t \simeq 1$~ps  is estimated from the short MD simulations run during the iMapD exploration, measuring the time it takes the system to travel a distance $\sim\sigma$.

\begin{figure}[t!]
	\includegraphics[width=\textwidth]{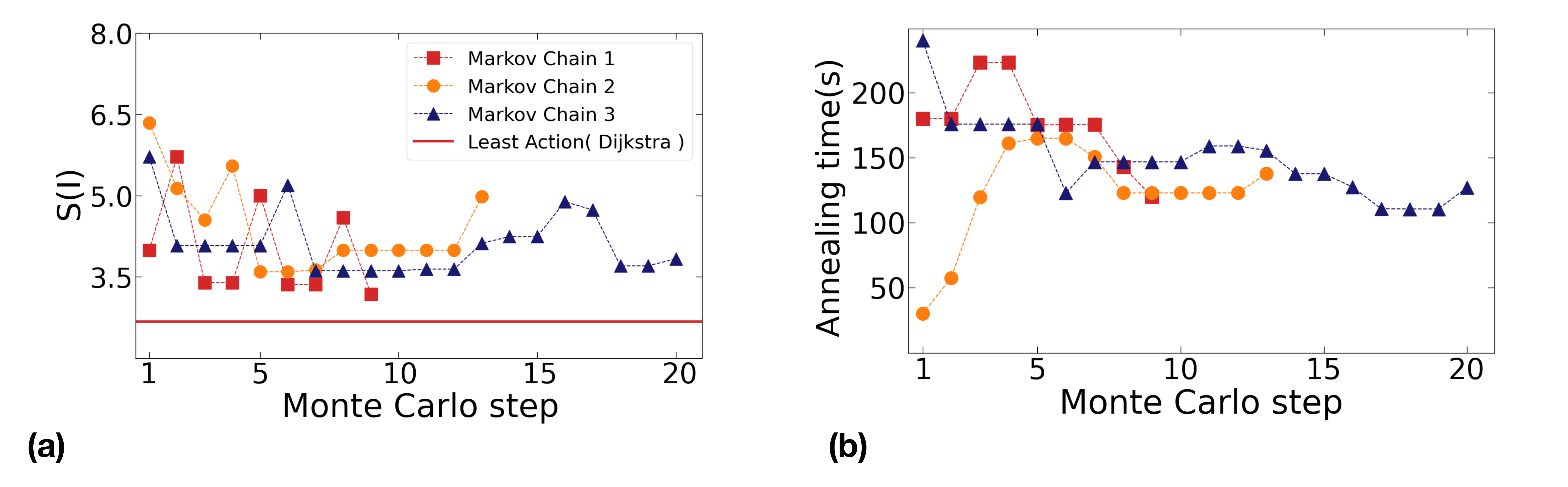}
	\caption{ Evolution of the path action $S(\I)$ (a) and annealing time $\ts$ (b) along the Monte Carlo paths generated using the hybrid classical/quantum annealing implemented on D-Wave.}
		\label{Results_1}
\end{figure}

 To implement our hybrid classical/quantum Monte Carlo scheme, we encode the quantum Hamiltonian $\hat H$ defined by the graph using the Ocean suite, operating on the D-Wave. Encoding our system on D-Wave requires 578 qbits, given by the sum  of the number of nodes and edges of our network. To generate trial paths, we rely on the hybrid solver available on Leap, which combines quantum annealing with classical simulating annealing. In this case, $\ts$ is identified with the total quantum and classical computing time employed by the solver. 
We estimate the resulting conditional probability $P(\I|\ts)$ entering Eq.~(\ref{detbal})  by means of a direct calculation on D-Wave (\ref{table:table1}), using Eq.~(\ref{cumul}). In Fig.~\ref{Pcondest},  we report the average value of the energy $\overline{E}$ and its standard deviation $\Delta$, entering  Eq.~(\ref{cumul}).

We initiated three independent Markov chains from  arbitrary paths generated by a quantum annealing process at $\ts=180$~s, $30$~s, and $240$~s, corresponding to about $8.6$~s, $1.4$~s and $11.4$~s of quantum annealing time, respectively (details on how we determine the initial and final nodes are outlined in the Appendix~\ref{appendixd}). We evolved $\ts$ according to Eq.~(\ref{evolts}) with  $k=2\times10^{-4}$~s$^{-1}$ and $t_0=150$s and then accepted or rejected the new paths according to Eq.~(\ref{detbal}). The parameter determining the relative strength of the constraint and target Hamiltonians was set to $\alpha=\sum_{ij}w_{ij}$. With this choice, on average, over $60\%$ of the annealing sweeps led to configurations of binary variables $\Gamma^{(1)}$ and $\Gamma^{(2)}$ with a correct path topology (\ref{table:table2}), thus providing viable trial transition paths.

In Fig.~\ref{Results_1}, we show the change in path action $S$ (left panel) and the hybrid minimization time $\ts$ (right panel), along our three Markov chains. As these results show, the Monte Carlo algorithm occasionally accepts trial moves with a higher action. They also show that longer annealing times do not always yield paths with  lower actions. This is expected,  since the $P(E|\ts)$ distributions have significant overlap, as it can be inferred from Fig.~\ref{Pcondest}.

The transition paths generated by our TPS scheme are consistent with the free energy landscape produced by equilibrium MD. Figure~\ref{Results_3}(a) shows the first and last accepted transition paths of one of the generated Markov chains. Both paths correctly connect the two meta-stable states, navigate the low-free energy regions of the surface, and cross the barrier at its lowest point. 

The transition paths explore a region around the most probable path (\cite{DRP1, DRP2, DRP0}), which in  Fig.~\ref{Results_3}(b) is shown based on calculations on a classical computer using the Dijkstra algorithm~\cite{Dijkstra}. Figure~\ref{Results_3}(b) also reports how often the sampled transition paths pass the nodes of the network, i.e., the statistical weight of the corresponding configuration in the transition path ensemble. All transition paths go through the transition state. However, due to the presence of fluctuations, a finite probability is obtained also at configurations with relatively high free energy. The deterministic Dijkstra algorithm can only detect the global minimum of the functional $S(\I)$. In contrast, our TPS algorithm accounts for fluctuations that lead to the full transition path ensemble.

\begin{figure}[t!]
	\includegraphics[width=.85\textwidth]{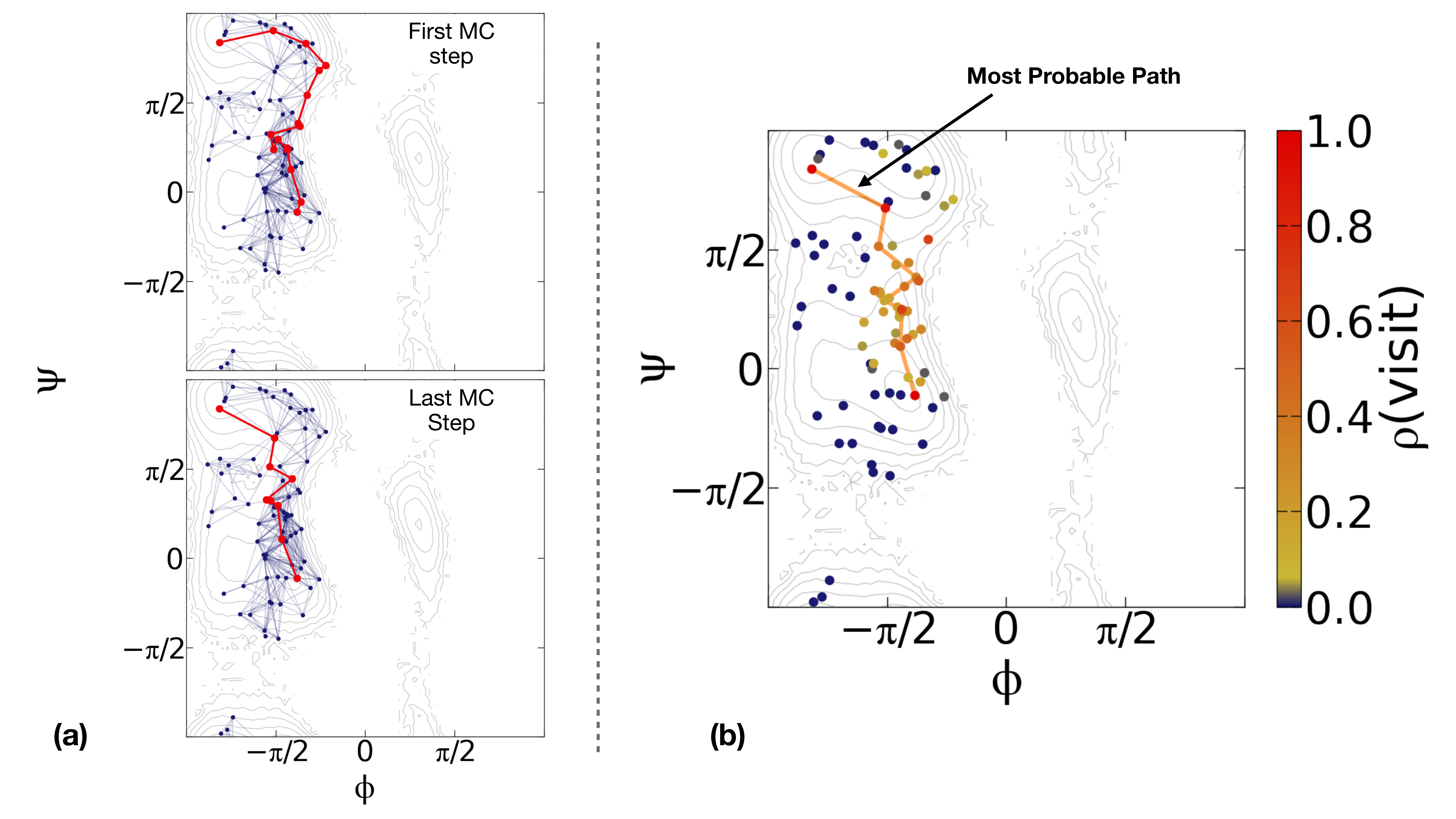}
	\caption{(a) Transition pathways for the $C_5\to\alpha_R$ transition of alanine dipeptide obtained from our Monte Carlo scheme. The red line denotes the first (top) and last (bottom) trajectory in  a Markov chain. The points in the Ramachandran plots are obtained from projecting the configurations generated with iMapD. In the background is the free energy surface calculated from 1 $\mu$~s of plain MD. (b) Transition path density on the Ramachandran plane, evaluated for the ensemble of trajectories calculated with our Monte Carlo scheme. The solid orange line is the most probable path, obtained using the Dijkstra algorithm~\cite{Dijkstra} on a classical computer. }
		\label{Results_3}
\end{figure} 

The main advantage of our hybrid classical/quantum scheme is that it allows us to efficiently obtain independent  transition paths. Indeed, we expect that the only source of correlation should be that introduced by the Markovian stochastic evolution of the minimization time $\ts$ (Eq.~(\ref{evolts})).
To quantify the degree of decorrelation in the ensemble of  trajectories sampled in three Markov chains, we consider the auto-correlation function
\be
G(N) = \frac{1}{N_{\mathrm{MC}}}\sum_{k=1}^{N_{\mathrm{MC}}} \left[\left\langle \frac{1}{|\mathcal{E}|}\sum_{i<j}'\Gamma_{ij}^{(2)}(k+N)\Gamma^{(2)}_{ij}(k)\right\rangle -  \left\langle \frac{1}{|\mathcal{E}|}\sum_{i<j}'\Gamma_{ij}^{(2)}(k+N)\right\rangle \left\langle \frac{1}{|\mathcal{E}|}\sum_{i<j}'\Gamma_{ij}^{(2)}(k)\right\rangle\right].
\label{GNeq}
\ee
In this equation, the $\Gamma^{(2)}(k)$ represent the collection of all  binary link  variables generated at the  $k$-th Monte Carlo step and we are implicitly assuming periodic boundary conditions, i.e., $\Gamma^{(2)}_{ij}(N_{\mathrm{MC}}+1) =\Gamma^{(2)}(1)_{ij}$. 
Here, $N_{\mathrm{MC}}$ is the number of Monte Carlo steps, $\sum'$ denotes a summation restricted to the $\epsilon$ pairs of indexes $i$ and $j$ that label topologically adjacent sites in the graph and $|\mathcal{E}|$ denotes the number of edges present in the graph.
Finally, $N$ is the distance in the Monte Carlo chain.
The average $\langle\cdot\rangle$ is intended over many Monte Carlo trajectories.
In practice, however, the computing time that was available to us on the D-Wave quantum computer was sufficient to generate only 3 Monte Carlo trajectories. Since  such a limited statistics does not allow us to estimate the averages in Eq.~(\ref{GNeq}), 
in Fig.~\ref{GN} we plot the behaviour of $G(N)$ (evaluated relatively to its initial value $G(0)$) for each independent Markov chain. These results clearly indicate that the correlation in the link binary variables is strongly suppressed after just a single Monte Carlo step.

\section{Conclusions}
\label{conclusions}
In this work, we have established a novel computational framework to sample the transition path ensemble of molecular conformational transitions, which integrates a  ML driven exploration with a hybrid Monte Carlo scheme that exploits the potential of  QC. 
We have used the iMapD algorithm~\cite{IMapD} to achieve an uncharted exploration of the molecular intrinsic manifold, without introducing any choice of CV, nor biasing force. These data enabled us to build a coarse-grained representation of the dynamics directly on the  intrinsic manifold. To construct this low-resolution theory, we adapted regularization and renormalization methods that were originally developed within the context of high energy physics~\cite{Howto}, and which may also be useful in other applications in soft-condensed matter and biophysics~\cite{Corradini, MSMRG}. 
We then encoded the path sampling problem in a form that enabled us to use a D-Wave quantum annealer to generate uncorrelated trial transition paths, thus enhancing the exploration of the transition path ensemble. Finally, by using the Metropolis criterion in Eq.~(\ref{detbal}), we made sure to account for each trajectory with its correct statistical weight in the transition path ensemble. 

The algorithm we presented here is designed to sample the full transition path ensemble. This achievement represents a crucial advancement with respect to previous attempts to compute the most probable transition paths on a quantum computer~\cite{QDRP}.
The transition path ensemble is often  heterogeneous,  displaying several alternative transition channels, corresponding to alternative molecular mechanisms. Even though in the proof-of-concept we discussed here we restricted our sampling to 23 transition paths, in general the number of trajectories is only limited by the available computational resources.

While significant effort has has been made towards designing quantum algorithms for quantum many-body problems~\cite{Q1,Q2, Q3,Q4, Q5, Q6, Q7, Q8, Q9, Q10, Q11}, only a few applications of quantum computing to classical molecular sampling problems have been reported to date~\cite{QS1, QS2, QDRP,POLYQ, Mazzola2021}. Most of the these attempts assume a  simplified  molecular representation, among which lattice discretization~\cite{QS1, QS2, POLYQ}. 
Unlike the method developed in~\cite{QDRP}, which was designed to return only the most probable path, here we sample the full transition path ensemble. In addition, we do not introduce any unphysical biasing force nor a choice of CVs to accelerate the exploration of configuration space. Finally, to the best of our knowledge, the present calculation represents the first successful application of a quantum computing machine to characterize a molecular transition using a state-of-the-art atomistic force field.

\begin{figure}[t!]
	\includegraphics[width=0.6\textwidth]{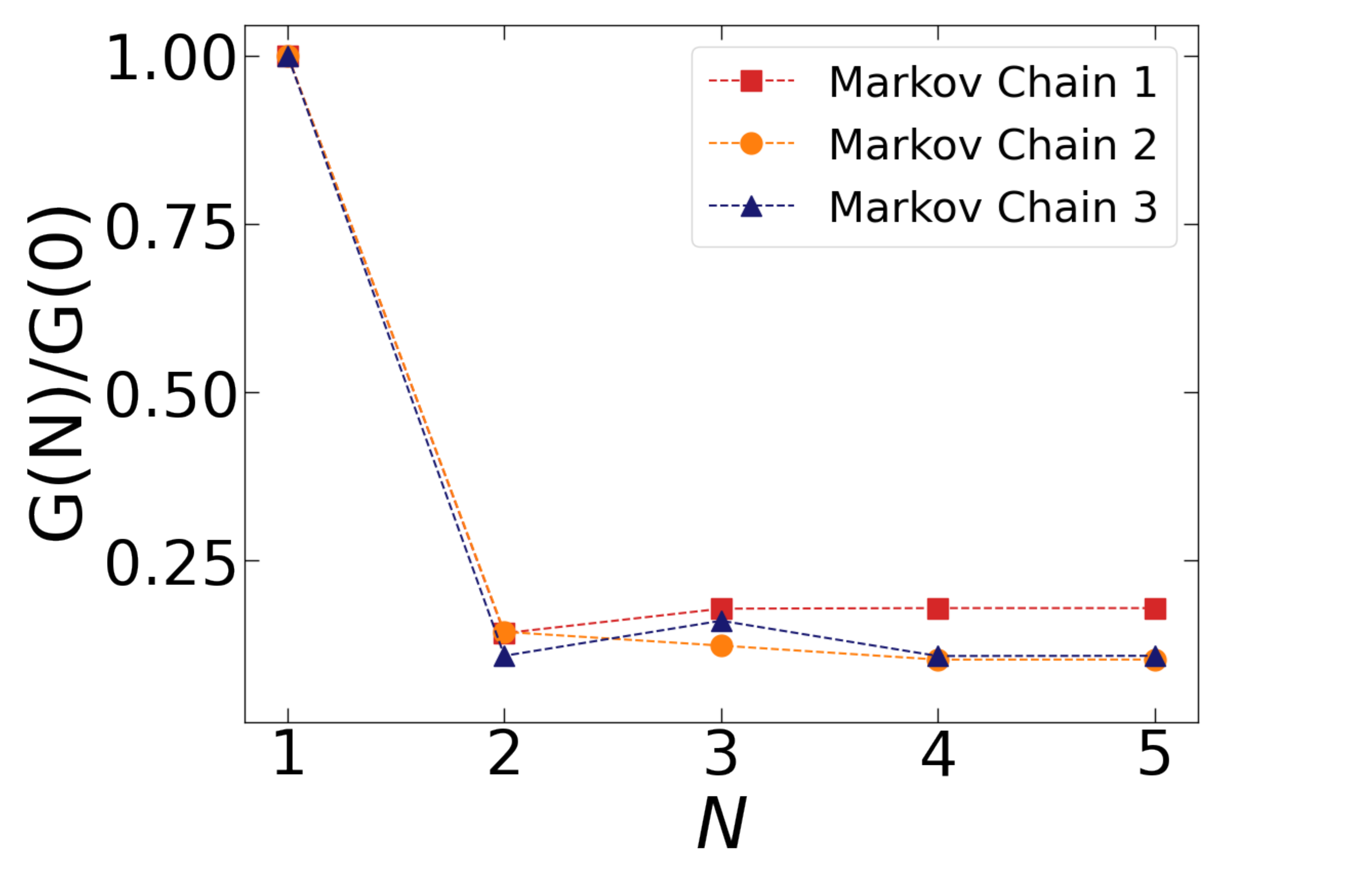}
	\caption{The ratio of auto-correlation function $G(N)/G(0)$ (see Eq.~(\ref{GNeq})) plotted as a function of Monte Carlo steps $N$ for three independent Markov chains.\label{GN} }
\end{figure}

With the present quantum encoding, the size of the molecular systems that  can presently be investigated is  limited by the relatively small  number of qubits that are available on the existing quantum annealing machines. The characterization of  transitions with a comparable level of spatio-temporal resolution of much larger molecules (for example, the folding of a small protein)  typically requires to generate  $10^3-10^4$ points on the intrinsic manifold~\cite{QDRP}.
In this case, implementing our scheme on a quantum computer would require a number of qubits more than one order of magnitude larger than that of the most powerful existing quantum annealing device.  However, if the size of quantum computing hardware continues to  grow in size and performance according to the present exponential rate~\cite{Scholl_2021,Pogorelov_2021,Ball_2021}, we may hope this threshold to be reached within the foreseeable future.
Once quantum annealing machines with order of $10^5$ qubits become available, we envision our scheme to provide a powerful new paradigm for simulating complex molecular transitions without any need of prior knowledge, with great potential for applications in biophysics and material science.

\acknowledgments
We acknowledge  important discussions with L.\ Tubiana and A. Roggero. R.C.\ acknowledges the support of the Frankfurt Institute for Advanced Studies and the LOEWE Center for Multiscale Modelling in Life Sciences of the state of Hesse. 
P.H. acknowledges support by the ERC Starting Grant StrEnQTh (project ID 804305), the Google Research Scholar Award ProGauge, Provincia Autonoma di Trento, and Q@TN — Quantum Science and Technology in Trento.  

\appendix
\section{Details of the the intrinsic manifold exploration}\label{appendixa}
We start by performing short unbiased sampling in each meta-stable state. Here, we assume that the interesting meta-stable states are known and that representative structures are available, as it is often the case. 
Each local sampling returns a set of $M$ configurations $\mathcal{C}=\left\{ Q_1,\dots, Q_M \right\}$, with $\mathrm{dim}(Q_i)=3N_a$. We identify the boundary of $\mathcal{C}$ in the low-dimensional intrinsic manifold defined by  diffusion maps~\cite{DMAP}. For each couple of configurations $Q_i$ and $Q_j$, we calculate the pairwise root-mean-square-deviation (RMSD) on all non-hydrogen atoms after removing global translations and rotations, and use it as a metric, $d_{ij} = \mathrm{min\,RMSD}(Q_i, Q_j)$. We then construct a transition matrix by calculating
\begin{align}
    P_{ij} = C~e^{-d_{ij}^2/\epsilon^2}\,,
    \label{EQ:DMAP_probability}
\end{align}
where $\epsilon$ is a distance threshold that defines a notion of neighbourhood in the configuration space, here defined as $\epsilon=\sigma(d_{ij}) -\Delta\sigma(d_{ij})$, where $\sigma$ is the average of the pairwise distances, and $\Delta\sigma$ its standard deviation. We use here a Guassian kernel, which is a popular choice, but other choices are possible~\cite{DMAP}, $C$ provides a normalization such that the sum along each row of $P_{ij}$ is one. Solving the eigenvalue problem for $P_{ij}$, we obtain eigenvalues $\lambda_k$ and eigenvectors $\psi_k$, where $k=0,\dots, M-1$. A projection of the high-dimensional data $\mathcal{C}$
onto a $n \ll N$ low-dimensional embedding $\mathbf{z} = \{z_1, ..., z_M\}$ is established via the components of $n$ dominant eigenvectors~\cite{DMAP}, i.e., each $Q_i$ gets mapped into $z_i = \{\psi_1(i), \dots \psi_n(i)\}$, where $\psi_k(i)$ is the $i$-th component of eigenvector $\psi_k$.  
We identify points on the boundary of $\mathcal{C}$, the set of which we denote $\mathcal{C}^B$, as those defining a convex hull containing the low-dimensional embedding $\mathbf{z}$.

We then generate new configurations in the unexplored regions by ``shooting" beyond the boundary of the known configurations space. For each point $Q_i^B$ in $\mathcal{C}^B$, we first identify the set of nearest neighbours within a given distance $\delta=\sigma(d_{ij}) -\Delta\sigma(d_{ij})$. For each such set, we calculate the projections of all points in the principal component analysis (PCA) representation, i.e., $ q_i = Q_i\mathbf{V}$, where lower case indicates a sample projected on the principal components, upper case in the configuration space, and $\mathbf{V}$ is a matrix containing the PCA loadings~\cite{IMapD}. For every point on the boundary and its neighborhood, we generate a new unexplored configuration beyond the boundary with 
\begin{equation}
    q_{\mathrm{new}} = q_{{\mathrm{b}}}-q_{c} + c \frac{q_{\mathrm{b}}-q_{\mathrm{c}}}{|q_{\mathrm{b}}-q_{\mathrm{c}}|}.
\end{equation}
Here, $q_{\mathrm{new}}$, $q_{\mathrm{b}}$, and $q_{\mathrm{c}}$ are respectively coordinates in the principal component projection for the new data point, boundary point, and the center of mass of the neighboring set without including the boundary point. The constant $c > 0$, which is adjusted heuristically, controls how far away beyond the boundary we generate new configurations. Finally, we retrieve the Cartesian coordinates of the new configuration with $Q_{\mathrm{new}}=q_{\mathrm{new}}\mathbf{V}^T+Q_{c}$.

The exploration proceeds iterating between two steps: generating new configurations beyond the boundary and starting short rounds of unbiased sampling from these configurations. At every step $i$, we generate one new configuration for every point on the boundary of the data set of all configurations sampled up to step $i-1$. We then merge together all configurations sampled up to step $i$ and identify a new boundary, which is then used in the $i+1$-th iteration.

\section{Regularization of $\langle Q_{n+1}|  e^{-\frac{1}{\hbar_{\textrm{eff}}} \hat H_{\textrm{eff}} \Delta t}|Q_{n}\rangle$}
In this Appendix, we  derive the regularized expression for the elementary propagator.
Following the standard derivation of Feynman's path-integral, we first split kinetic and potential contributions:
\begin{eqnarray}
&&\langle Q_{n+1} | e^{-\frac{\Delta t}{\hbar_{\textrm{eff}}} \hat H_{\textrm{eff}}}|Q_n\rangle_{\textrm{reg}} = \int dz \langle Q_{n+1} | e^{-\frac{\Delta t}{\hbar_{\textrm{eff}}} \hat T_{\textrm{eff}}}| z\rangle_{\textrm{reg}} \langle z | e^{-\frac{\Delta t}{\hbar_{\textrm{eff}}} \hat V_{\textrm{eff}}}| Q_n\rangle_{\textrm{reg}}\,,
\end{eqnarray}
where we inserted the regularized resolution of the identity  $1 = \int dz |z\rangle_{\textrm{reg}}\langle z |$. 
Let us first compute the kinetic energy term:
\begin{equation}
\langle Q_{n+1} | e^{-\frac{\Delta t}{\hbar_{\textrm{eff}}} \hat T_{\textrm{eff}}}| z\rangle_{\textrm{reg}} = 
\dep  e^{\frac{i}{\hbar_{\textrm{eff}}} P (x_{i+n}-z)} e^{- \frac{P^2 \Delta t}{2 m \hbar_{\textrm{eff}}} \left(1 +  \frac{m \sigma^2}{ \Delta t\hbar_{\textrm{eff}}}\right)}= \mathcal{N} e^{-\alpha\frac{m(Q_{n+1}-z)^2}{2  \hbar_{\textrm{eff}} \Delta t}  },
\end{equation}
where $\alpha = \frac{1}{\left(1 +  \frac{m \sigma^2}{ \Delta t\hbar_{\textrm{eff}}}\right)}$ and $\mathcal{N}$ is an irrelevant normalization constant. 

We now discuss the regularized expression for the term containing the effective potential. Assuming sufficiently small $\Delta t$, we can evaluate the matrix element  $\langle z|  e^{-\frac{\Delta t}{\hbar_{\textrm{eff}}} \hat V_{\textrm{eff}}(Q)} | Q_n \rangle_{\textrm{reg}}$ by expanding to leading order the exponent. A straightforward calculation yields:
\be
\langle z |  e^{-\frac{\Delta t}{\hbar_{\textrm{eff}}} \hat V_{\textrm{eff}}} | Q_n \rangle_{\textrm{reg}}
&\simeq& e^{-\frac{\Delta t}{\hbar_{\textrm{eff}}} V^{\textrm{reg}}_{\textrm{eff}}(Q_n)}\delta_\sigma(z-Q_n),
\ee
where
\be\label{veffR}
V^{\textrm{reg}}_{\textrm{eff}}(Q) =  \dep 
~e^{\frac{i}{\hbar_{\textrm{eff}}}P\cdot  Q} 
~e^{-\frac{P^2 \sigma^2}{2 \hbar_{\textrm{eff}}^2}}~\tilde{V}_{\textrm{eff}}(P)
\ee
is the regularized effective potential.

\section{Fast exploration of alanine dipeptide's configuration space}
\label{appendixc}
\subsection{Exploring the intrinsic manifold}

We start by sampling from two metastable states of alanine dipeptide in OpenMM. One starting from $C_5$ region at the top left corner of Fig.~\ref{fig:FEL}, and from $\alpha_R$ in the vicinity of $(\phi=-75$ ,$\psi=-20)$.
\begin{figure}[ht]
 \centering
 \includegraphics[width=.5\textwidth]{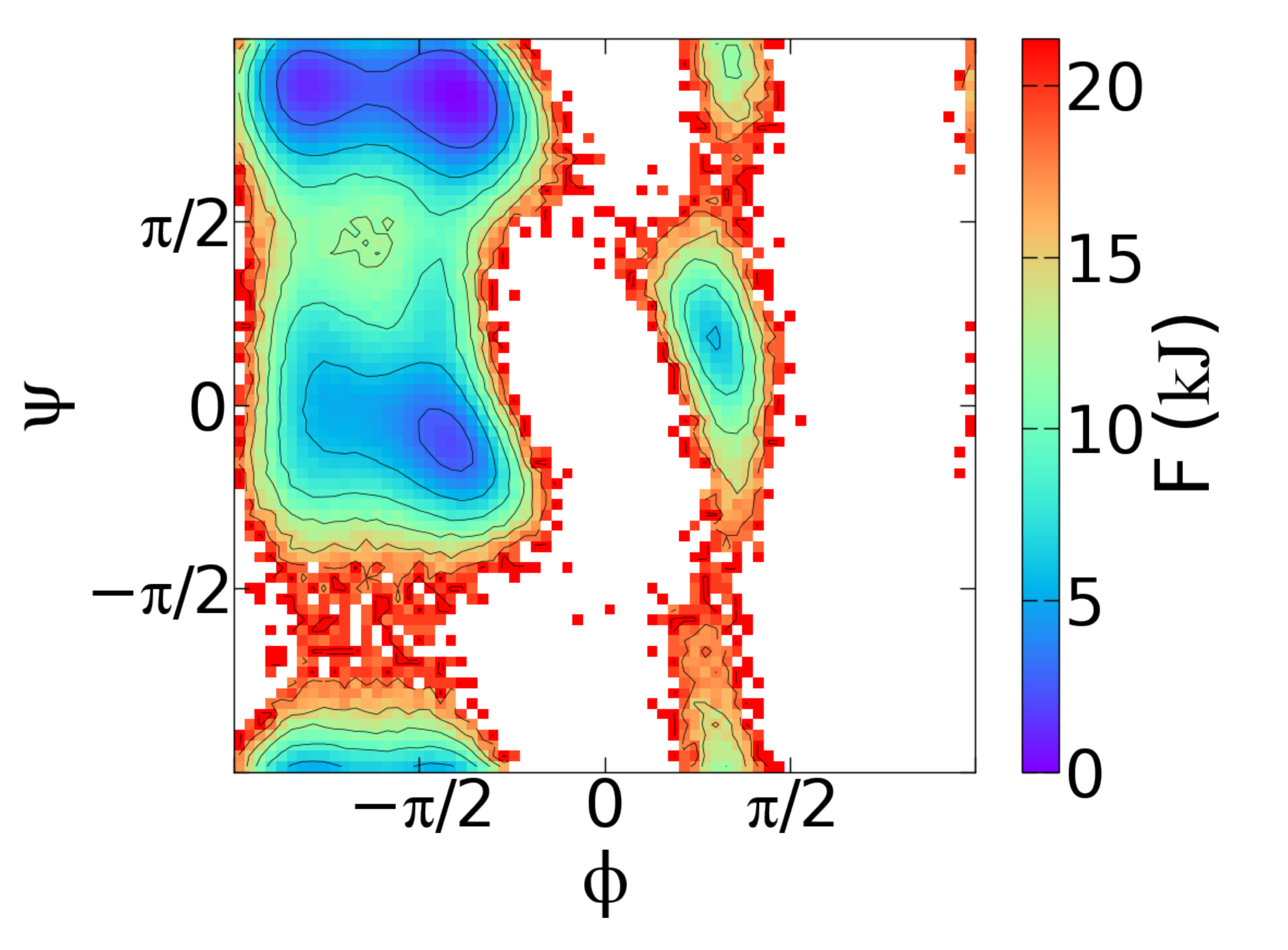}
 \caption{Free energy Landscape of alanine dipeptide, projected onto its two main dihedral angles. This figure was generated by simulating alanine dipeptide for $1~\mathrm{\mu s}$ at $T=300~$K, and in explicit TIP3P water. Contour lines are drawn every 3 kJ.
 \label{fig:FEL}} 
\end{figure}
After this initial sampling, we evaluate the diffusion map (DMAP) for each set of configurations separately to obtain a low-dimensional representation of the sampled configurations . We then identify the boundary in this representation, to initialize new simulations beyond the region that has already been sampled. We measure the pairwise distance between configurations by using the RMSD calculated on all non-hydrogen atoms after having removed global translations and rotations.
The exploration proceeds by initiating unbiased sampling from each new configurations and then merging all the new data to the previous. By iterating over DMAP evaluation at every step, finding new configurations, and sampling we populate the transition region between $C_5$ and $\alpha_R$. 

We eventually terminate the iterations when the configurations explored starting from the two initial metastable state overlap, i.e. when at least two configurations have RMSD closer than 0.3 nm (Fig.~\ref{fig:final_data_ALA}(a)).

\begin{figure}[ht]
 \centering
 \includegraphics[width=\textwidth]{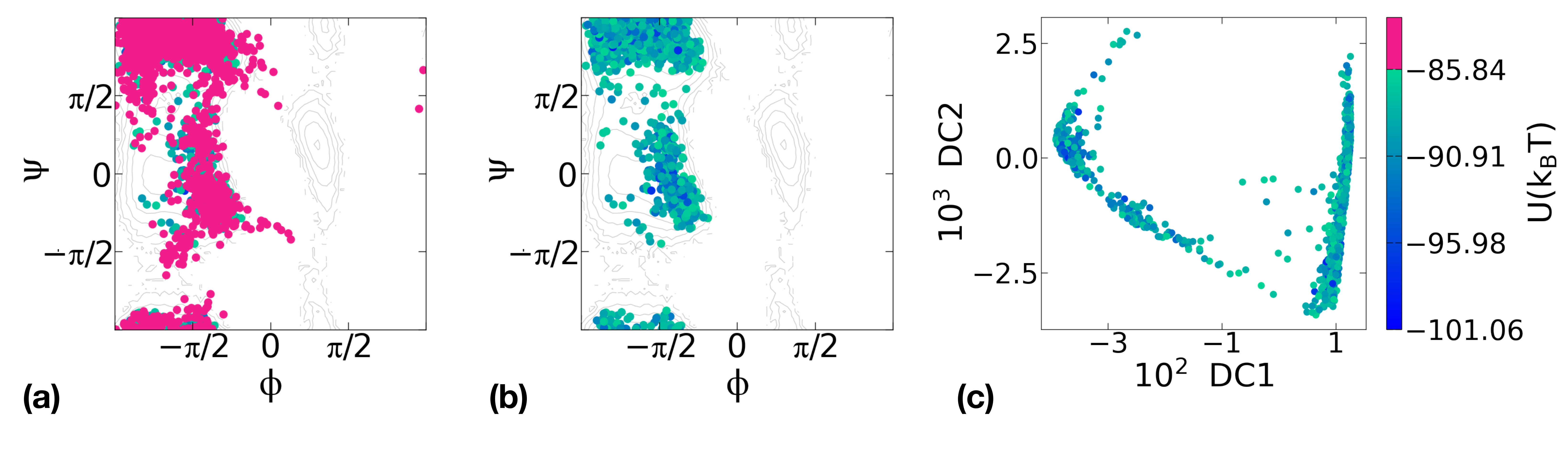}
 \caption{a) Final result of the exploration with iMapD. b) We further remove those points with higher potential energy than the median ($\approx\;85.84~k_\mathrm{B}T$), shown as dark pink color. c) DMAP embedding of points whose potential is below median, shown as a function of the first two diffusion coordinates (DCs).\label{fig:final_data_ALA}} 
\end{figure}
\subsection{Simulation details}
In all simulations, the molecule was placed in a square box with 2.85 nm base vector, solvated with the TIP3P water model using AMBER99SB forcefield \cite{forcefield}. We energy-minimized the initial configuration using L-BFGS algorithm implemented in OpenMM \cite{OpenMM}, with tolerance of 500 kJ. Simulations were performed at a temperature $T=300~$K, using a Langevin integrator with friction coefficient $\gamma=91~\mathrm{ps^{-1}}$ and timesteps of $\Delta t=2$ fs. The initial sampling bursts in the two metastable states were $200$ ps long starting in $C_5$ state and $20$ ps for $\alpha_R$. All following runs were $1$ ps long.

\section{Graph Representation of the coarse-grained dynamics for alanine dipeptide\label{appendixd}}

\subsection{Data Reduction and Identification of the Nodes in the Graph}
Successive rounds of explorations carpeted the transition region between the basins associated with the $\alpha_R$ and $C_5$ states. The number of sampled configurations needed to be reduced to obtain a sparser transition network compatible with the D-Wave quantum annealer. We first removed all configurations having a potential energy higher than the median value $85.84~k_{\mathrm{B}}T$ (Fig.~\ref{fig:final_data_ALA}(b)). Secondly, we calculated the DMAP of all remaining samples and projected them on the first two diffusion coordinates (DCs), obtaining a set of points $\mathcal{C}=\{z_i\}$ (Fig.~\ref{fig:final_data_ALA}(c)).
We then proceeded by identifying the two configurations in $\mathcal{C}$ with minimum RMSD distance from the initial configurations lying in two basins $C_5$ (denoted by $Q_A$) and $\alpha$ (denoted by $Q_B$). Starting from $Q_A$ ($z_A$ in DCs), we kept
the nearest point $z_{k1}$ satisfying $D_{\mathrm{diff}}(z_{k1},z_A)>D^{\mathrm{thresh}}_{\mathrm{diff}}$, and removed all the configurations with lower $D_{\mathrm{diff}}$ from $\mathcal{C}$. Here $D_{\mathrm{diff}}$ (diffusion distance) is the Euclidean distance between points in DMAP space, defined here as the planed spanned by the two diffusion coordinates $DC1$ and $DC2$. We used $D^{\mathrm{thresh}}_{\mathrm{diff}}=7.5\times 10^{-4}$ inspecting the histogram of nearest-neighbour pairwise diffusion distances among the points in $\mathcal{C}$, Fig~\ref{fig:trimmed_data}(a). This value ensures that for any configuration in the original data set only one representative is kept, and henceforth their nearest neighbors based on $D_{\mathrm{diff}}$---considered kinetically similar---are removed. 
We continued by applying the same procedure between remaining configurations in $\mathcal{C}$ and $z_{k1}$, and therefore identifying $z_{k2}$. We iterated over this step until there was no remaining element in $\mathcal{C}$. Finally, we appended back $z_B$ to $\mathcal{C}$ (in case it was removed in previous iterations) to obtain a reduced version of our data set containing $83$ points that approximately span uniformly the DMAP embedding of the transition between the two metastable states (Fig.~\ref{fig:trimmed_data}(b)-(c)).

\begin{figure}[ht]
\centering
 \begin{subfigure}{\textwidth}
 \includegraphics[width=\textwidth]{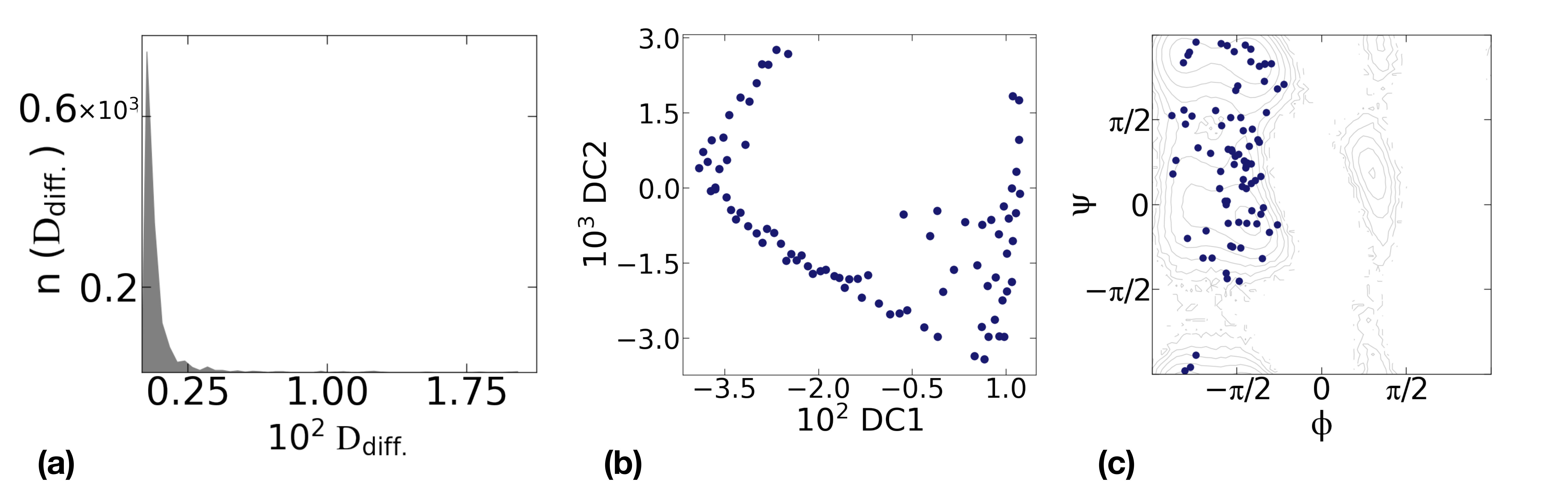}
 \end{subfigure}
 \caption{Sub-sampling explored configurations in a kinetically uniform way a)$D_{\mathrm{diff}}$ histogram for nearest neighbors. $D^{\mathrm{thresh}}_{\mathrm{diff}}=7.5\times 10^{-4}$ contains $>98\%$ of nearest neighbor pairwise distances. Reduced data projected on b) the first two DCs and c) Ramachandran plot.}
 \label{fig:trimmed_data}
\end{figure}

\subsection{Building the Graph}
The next step required to build a graph having as nodes the sparse set of configurations $\mathcal{C}_{reduced}$. In this graph, only nodes that correspond to configurations that are kinetically and structurally close should be connected. We used two criteria to insure this condition. Two nodes should be connected if their diffusion distance is smaller than 0.01 and their RMSD closer than 0.5 nm. Both thresholds were chosen heuristically from the histogram of pairwise diffusion distances and RMSD calculated on the set of configurations generated by iMapD (Fig.~\ref{fig:density_plot}).

\begin{figure}[ht]
\centering
 \begin{subfigure}{\textwidth}
 \includegraphics[width=\textwidth]{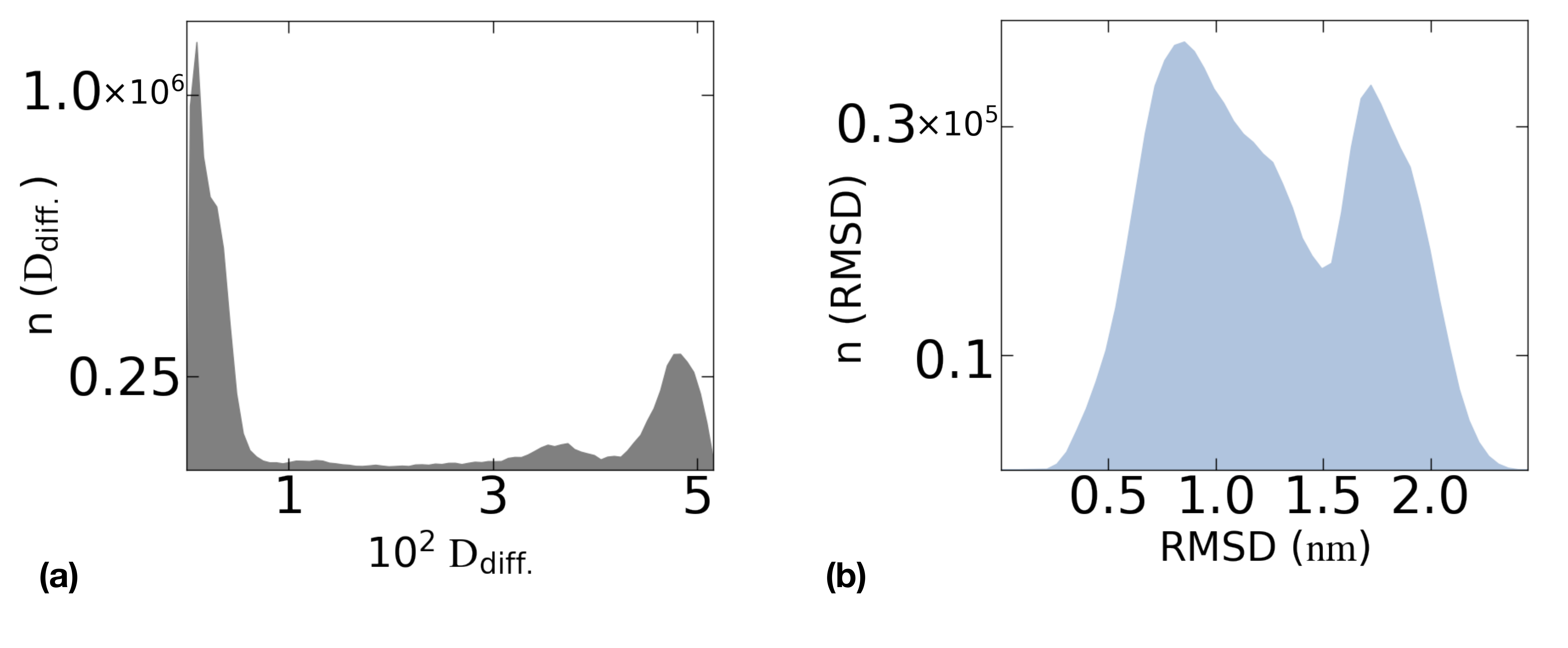}
 \end{subfigure}
 \caption{a) Histogram of pairwise $D_{\mathrm{diff}}$ b) and of pairwise RMSD calculated on all configurations sampled by iMapD.\label{fig:density_plot} }
\end{figure}

\subsection{Details on the Calculation of the Graph Weights $w_{ij}$}

Eq.~(\ref{wijdef}) provides a suitable definition of the graph weights $w_{ij}$ that explicitly depends on the time resolution $\Delta t$ of our coarse-grained representation of the dynamics. 

An alternative way of fixing a time resolution is to consider the Laplace transform of the elementary propagator, 
\begin{eqnarray}\label{Lap}
G(Q_f, s|Q_i) = \int_0^\infty d\tau e^{-s \tau} K(Q_f, \tau|Q_i,0).
\end{eqnarray}
In this expression, the time resolution is determined by the inverse of the frequency cut-off scale $s$. 
We adopted this regularization prescription in our illustrative application to alanine dipeptide. 

To evaluate the elementary propagator in Laplace space, we resorted to the so-called Dominant Reaction Pathway (DRP) formalism \cite{DRP1,DRP2}, that originates from analyzing the path integral expression of the Feynman propagator $K(Q_f,t| Q_i,0)$ in saddle-point approximation.

The saddle-point condition yields a Newton-type equation of motion for the most statistically relevant trajectory $\bar Q(\tau)$,
\begin{equation}\label{Newton}
m \ddot{\bar{Q}} = \nabla V_{\textrm{eff}}(\bar{Q}),
\end{equation}
where $\overline{Q}(\tau)$ obeys the boundary conditions $\bar Q(t)=Q_j$ and $\bar Q(0)= Q_i$.

Eq.~(\ref{Newton}) implies the conservation of the effective energy 
\be
E_{\textrm{eff}}= \frac{\dot Q^2}{2} m -V^{reg}_{\textrm{eff}}(Q)
\ee
along the most probable path, $\bar{Q}(\tau)$.

Using the conservation of $E_{\textrm{eff}}$ and recalling the definition $\hbar_{\textrm{eff}}=2 \frac{k_BT}{\gamma}$ , the Feynman path integral is expressed as
\begin{eqnarray}\label{DRPP}
P(Q_f, t|Q_i) \simeq \mathcal{N}~e^{-\frac{1}{2 k_BT} \left( U(Q_f)-U(Q_i)\right)}~e^{\frac{\gamma}{2 k_B T}\left(E_{\textrm{eff}} t - S_{HJ}[\bar Q]\right)},
\end{eqnarray}
where $\mathcal{N}$ is an irrelevant normalization factor, $S_{HJ}[\bar Q]$ is the so-called Hamilton-Jacobi (HJ) functional, 
\begin{equation}
 S_{HJ}[Q] = \int_{Q_i}^{Q_f} dl \sqrt{2 m\left(E_{\textrm{eff}} + V_{\textrm{eff}}[\bar Q(l)]\right)}, \end{equation}
and $dl$ is the Euclidean distance travelled along the saddle-point trajectory $\bar{Q}$.

Using expression, Eq.~(\ref{DRPP}), and recalling that $s t\gg 1$, the Laplace time integral, Eq.~(\ref{Lap}), finally becomes:
\begin{equation}\label{GE}
 G(Q_f, s| Q_i) = \mathcal{N} e^{-\frac{1}{2 k_BT} \left( U(Q_f)-U(Q_i)\right)}~e^{-\frac{\gamma}{2 k_B T} \int_{Q_i}^{Q_f} dl \sqrt{2m (s + V_{\textrm{eff}}[\bar Q(l)])}}.
\end{equation}
We note that the exponential pre-factor can be ignored when comparing the probability density of transition paths connecting the same initial and final configuration. Therefore, in the graph representation, we can retain only the second exponent. 

Eq.~(\ref{GE}) can be used to obtain the weights of the links in our coarse-grained graph representation of the Langevin dynamics. To this goal, the effective potential should be regularized, i.e. replaced by $V_{\textrm{eff}}^{reg}(Q) = C_V V_{\textrm{eff}}^{s}(Q)$. 
We obtain
\begin{equation}\label{SHJ1}
w_{i j} = \frac{|Q_i -Q_j|}{2 \sqrt{D}} \left( L_i+L_j \right),
\quad L_i = \sqrt{V^{reg}_{\mathrm{eff}}(Q_i)+s_0}
\end{equation}
where $D=k_BT/(m \gamma)$ is the diffusion coefficient. 
This definition ensures that that the weighted sum $S(\I)=\sum_{n=1}^{N-1} w_{i_n, i_{n+1}}$ provides a discretized representation of the HJ functional along the path $\I=(i_1, \ldots, i_N)$ .

An advantage of the frequency regularization (relative to the conventional time regularization) is that the resulting expression for $w_{ij}$ depends on the Hamilton-Jacobi functional, which is defined in terms of structural distances $\Delta l$ between configurations. The. convenience of this discretization arises from the fact that there is no gap in the internal distance scales of molecular systems.

\begin{figure}[!ht]
\centering
 \includegraphics[width=\textwidth]{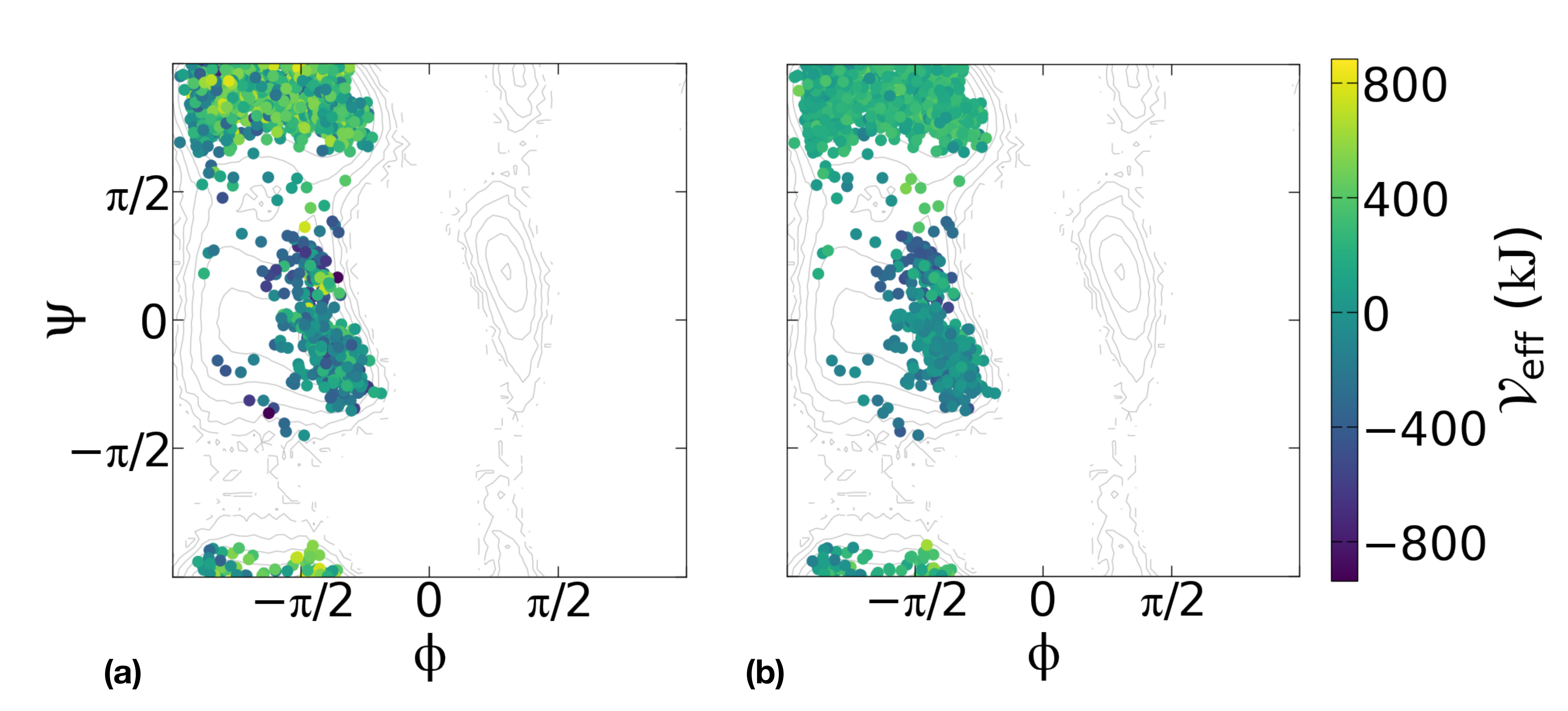}
 \caption{a) The original $V_{\textrm{eff}}$ and b) filtered version. In order to filter $V_{\textrm{eff}}$ after Fourier transforming its time-series, we introduced a symmetrical window on the frequencies that removed high frequency modes. We then performed an averaging procedure based on embedding of configurations on first two DCs. Therefore, we obtained a less fluctuating version of $V_eff$. \label{fig:effpot_filtered}}
 \label{}
\end{figure}

\begin{figure}[!ht]
\centering
 \includegraphics[width=\textwidth]{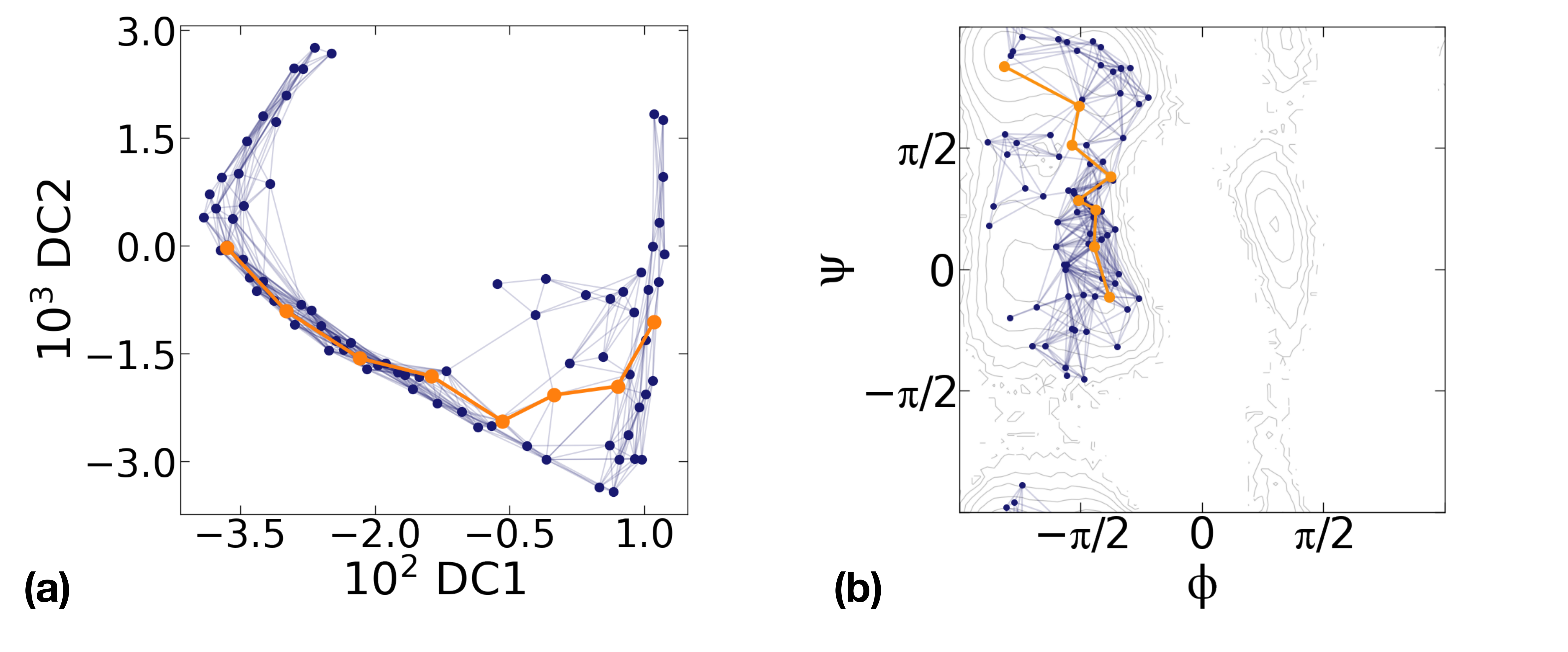}
 \caption{The network and most probable path, calculated via Dijkstra algorithm, in (a) DMAP embedding and (b) Ramachandran plot.\label{fig:dijkstra}}
\end{figure}

\begin{table}[!ht]
 \begin{tabular}{ | c | c | c | c |}
 \hline
 Num. of attempts & Correct topology & Wrong topology & Success rate\\ \hline
 117 & 69 & 48 & 0.59 \\\hline
 \end{tabular}
 \caption{Summary of the transition path generation on D-Wave to calculate the histogram reported in Fig.~\ref{Pcondest}. \label{table:table1}}
\end{table}

\begin{table}[!ht]
 \begin{tabular}{ | l | c | c | c | c |}
 \hline
 & Monte Carlo steps & Accepted paths & Wrong topology & Rejected paths \\ \hline
 Markov chain 1 & 9 & 7 & 0 & 2 \\ 
 Markov chain 2 & 13 & 8 & 2 & 3 \\ 
 Markov chain 3 & 20 & 10 & 4 & 6 \\
 \hline
 \end{tabular}
 \caption{Summary of the Markov chains sampling process on D-Wave.\label{table:table2}}
\end{table}

\end{document}